\documentclass[a4paper]{article}

\usepackage{a4wide}
\usepackage[UKenglish]{babel}
\usepackage{graphicx,pgfplots,subfigure}
	\pgfplotsset{compat=newest}
	\usepgfplotslibrary{fillbetween}
	\usepgfplotslibrary{groupplots}
	\usetikzlibrary{patterns}
	\usetikzlibrary{positioning}
	\usetikzlibrary{decorations.text}
	\graphicspath{{./}{figure/}}
\usepackage[colorlinks=true,linkcolor=blue,citecolor=red]{hyperref}
\usepackage{xcolor}
\usepackage{amsfonts,amsmath,amssymb,amsthm,mathrsfs,mathtools,bbold,empheq}
\usepackage[inline]{enumitem}
\usepackage{algorithm,algorithmic}

\newtheorem{theorem}{Theorem}[section]

\theoremstyle{remark}\newtheorem{remark}[theorem]{Remark}

\newcommand{\abs}[1]{\left\lvert#1\right\rvert}
\DeclarePairedDelimiter\ave{\langle}{\rangle}
\newcommand{\D}{\mathbb{D}}
\renewcommand{\div}{\operatorname{div}}
\newcommand{\cI}{\mathcal{I}}
\newcommand{\Lip}{\operatorname{Lip}}
\newcommand{\N}{\mathbb{N}}
\renewcommand{\P}{\mathbb{P}}
\newcommand{\pr}[1]{{}^\prime\!#1}
\newcommand{\R}{\mathbb{R}}
\newcommand{\supp}{\operatorname{supp}}
\newcommand{\cT}{\mathcal{T}}
\newcommand{\cU}{\mathcal{U}}
\newcommand{\V}{\mathbb{V}}

\allowdisplaybreaks

\begin{document}
\title{Markov jump processes and collision-like models \\ in the kinetic description of multi-agent systems}

\author{Nadia Loy\thanks{Department of Mathematical Sciences ``G. L. Lagrange'', Dipartimento di Eccellenza 2018-2022, Politecnico di Torino, Torino, Italy
			(\texttt{nadia.loy@polito.it})} \and
		Andrea Tosin\thanks{Department of Mathematical Sciences ``G. L. Lagrange'', Dipartimento di Eccellenza 2018-2022, Politecnico di Torino, Torino, Italy
			(\texttt{andrea.tosin@polito.it})}
		}
\date{}

\maketitle

\begin{abstract}
Multi-agent systems can be successfully described by kinetic models, which allow one to explore the large scale aggregate trends resulting from elementary microscopic interactions. The latter may be formalised as collision-like rules, in the spirit of the classical kinetic approach in gas dynamics, but also as Markov jump processes, which assume that every agent is stimulated by the other agents to change state according to a certain transition probability distribution. In this paper we establish a parallelism between these two descriptions, whereby we show how the understanding of the kinetic jump process models may be improved taking advantage of techniques typical of the collisional approach.

\medskip


\end{abstract}

\section{Introduction}
In recent times, the methods of kinetic theory have gained a lot of momentum in the modelling and the mathematical-physical understanding of multi-agent systems, i.e. systems composed by a large number of interacting individuals~\cite{pareschi2013BOOK}. The theoretical paradigm of the kinetic theory, rooted in the pioneering works by James C. Maxwell and Ludwig Boltzmann on gas dynamics~\cite{boltzmann1970CHAPTER}, allows one to investigate, from a truly multiscale perspective, the large scale collective trends emerging spontaneously from elementary microscopic interactions among the agents, by which the latter continuously change their microscopic state. Applications are nowadays variegated: an absolutely non-exhaustive list includes wealth distribution~\cite{cordier2005JSP}, opinion formation~\cite{boudin2009M2NA,toscani2006CMS,toscani2018PRE}, vehicular traffic~\cite{klar1997JSP,prigogine1971BOOK,puppo2017KRM}, crowd dynamics~\cite{agnelli2015M3AS,festa2018KRM}, biomathematics~\cite{hillen2006JMB,loy2019JMB}, up to control and uncertainty quantification problems~\cite{albi2016SIAP,albi2015CMS,albi2014PTRSA,duering2018EPJB,tosin2019MMS,tosin2019MCRF_preprint}.

The kinetic approach relies on a description of the microscopic dynamics as binary random processes. More specifically, the agents of the system are regarded as indistinguishable, therefore a prototypical interaction between two of them is assumed to be representative of the interaction of any randomly selected pair of agents. Typically, such a prototypical interaction is expressed by a \textit{collision-like model}, namely an explicit algebraic relationship between the pair of pre-interaction states of the interacting agents and the pair of their post-interaction states. This description is inspired by the well-known elastic collision rules of gas molecules, which prescribe that two particles, with respective pre-collisional velocities $v,\,v_\ast\in\R^3$, get the post-collisional velocities
$$ v'=v+[(v_\ast-v)\cdot n]n, \qquad v_\ast'=v_\ast+[(v-v_\ast)\cdot n]n, $$
$n\in\R^3$ being a unit vector in the relative direction of the centres of the two colliding particles. Another possible description of the microscopic dynamics, often employed in biomathematics for modelling particular types of cell motion~\cite{alt1980JMB,chauviere2007NHM,hillen2000SIAP,loy2019JMB,stroock1974ZWVG} but sometimes used also in different contexts~\cite{agnelli2015M3AS,bertotti2008MCM,delitala2014KRM,dolfin2017KRM,puppo2017CMS}, makes instead use of \textit{jump-like processes}. In this case, the possible post-interaction states of the agents are described by means of a transition probability distribution, thereby assimilating the microscopic dynamics to a (possibly continuous state) Markov-type process with state jumps as interaction outcomes. Both types of descriptions may be used as the basis for a mesoscopic representation of the multi-agent system, whereby one may deduce conceptually analogous but technically different kinetic models.

Collisional models are largely studied in the literature from both the analytical and the numerical point of view, see e.g.~\cite{pareschi2001ESAIMP,pareschi2013BOOK}. In particular, quite sophisticated tools have been developed to investigate and approximate the asymptotic distributions and the trend to equilibrium of these models, such as the \textit{quasi-invariant limit} and \textit{structure-preserving numerical schemes}~\cite{furioli2017M3AS,loy2019PREPRINT,pareschi2017JSC,toscani2006CMS}. On the contrary, for jump process models there are apparently much less general tools by which to tackle the explicit characterisation of the asymptotic distributions, which remains one of the major points in the study of kinetic models. Indeed, the asymptotic distributions depict the emerging behaviour, namely the observable manifestations, of the multi-agent systems at the aggregate level. Some results in this direction are available, however either generically in terms of existence and uniqueness of the equilibrium distributions under possibly restrictive assumptions on the admissible transition probabilities, see e.g.~\cite{freguglia2017CMS}, or only for linear kinetic models, i.e. those in which the agents do not interact with each other but with a fixed background, see e.g.~\cite{loy2019JMB}.

In this paper, we aim to establish a parallelism between the jump process and the collisional descriptions of the microscopic dynamics of a multi-agent system, by which to investigate, in a possibly approximate but explicit way, the large time trends of the former taking advantage of theoretical tools typically applied to the latter. More specifically, in Section~\ref{sect:micro_int} we review the microscopic and kinetic descriptions of either model and, in particular, we point out that a certain structure of the collisional model is able to reproduce exactly the time evolution of the mean and of the energy (i.e., the first two ``thermodynamical'' moments) predicted by the jump process model. By relying on this basic parallelism, in Section~\ref{sect:comparisons} we compare the time evolution of the kinetic distribution functions produced by either model in sufficiently simple but representative cases. We show that the transient regime is, in general, different but that, thanks to the same description of the first two thermodinamical moments, in some cases the asymptotic distributions, i.e. the ``Maxwellians'', actually coincide. Next, in Section~\ref{sect:quasi-invariant} we focus specifically on the explicit determination of the Maxwellian by considering, in particular, the so-called \textit{quasi-invariant interaction} regime. This is reminiscent of the \textit{grazing collision} regime, which, in the classical kinetic theory, has been introduced to approximate the large time trend of the collisional Boltzmann equation, hence in particular of its asymptotic distributions, by means of mean field-type Fokker-Planck equations~\cite{villani1998PhD,villani1998ARMA}. We generalise the definition of quasi-invariant interactions, so as to include also those described by jump process models. Then, thanks to the above parallelism, from the collisional description we determine explicitly some forms of the Maxwellian, which may consistently approximate the large time distribution of the jump process model. By means of analytical and numerical examples, we discuss the reliability of such an approximation with reference to applications such as cell motion and opinion dynamics. Finally, in Section~\ref{sect:bounded_domains} we indicate how to extend the theory to cases in which the microscopic states of the agents belong to bounded domains rather than to the whole real axis and in Section~\ref{sect:conclusions} we briefly outline some conclusions. The paper is completed by Appendix~\ref{app:MC}, in which we report, for completeness, the Monte Carlo algorithm that we have used to produce the numerical simulations of both the collisional and the jump process models.

\section{Microscopic interactions}
\label{sect:micro_int}
\subsection{Jump process-like models}
\label{sect:jump_process}
Let us consider a large system of indistinguishable agents characterised by a microscopic state $v\in\R$, which evolves in consequence of stochastic interactions. A probabilistic description of such interactions may be given as follows: if $\pr{v},\,\pr{v_\ast}\in\R$ are the \textit{pre-interaction} states of the interacting agents, we define a \textit{transition probability density} $T=T(v\,\vert\,\pr{v},\,\pr{v_\ast})$ such that the probability that the \textit{post-interaction} state $v$ of the first agent belongs to a certain subset $A\subseteq\R$ is
$$ \int_A T(v\,\vert\,\pr{v},\,\pr{v_\ast})\,dv. $$
Notice that $v\mapsto T(v\,\vert\,\pr{v},\,\pr{v_\ast})$ is a \textit{conditional} probability density, the pre-interaction states $\pr{v},\,\pr{v_\ast}$ playing the role of the conditioners. By choosing $A=\R$ we get, in particular,
$$ \int_\R T(v\,\vert\,\pr{v},\,\pr{v_\ast})\,dv=1, \quad \forall\,\pr{v},\,\pr{v_\ast}\in\R. $$
Such a microscopic description may be assimilated to a \textit{jump process}, $T(v\,\vert\,\pr{v},\,\pr{v_\ast})dv$ being the probability that an agent with state $\pr{v}$ jumps into $[v,\,v+dv]$ because of an interaction with an agent with state $\pr{v_\ast}$.

If $f=f(t,\,v):\R_+\times\R\to\R_+$ is the probability density function of the state $v$ at time $t$ in the considered population of agents, the kinetic model describing the evolution of $f$ under the interaction rules encoded in $T$ reads
\begin{equation}
	\partial_tf=\int_\R\int_\R T(v\,\vert\,\pr{v},\,\pr{v_\ast})f(t,\,\pr{v})f(t,\,\pr{v_\ast})\,d\pr{v}\,d\pr{v_\ast}-f.
	\label{eq:boltz.T}
\end{equation}
We will call~\eqref{eq:boltz.T} a \textit{jump process kinetic equation}. In addition to the stochastic microscopic dynamics described above, the basic ingredient leading to this equation is the so-called \textit{Boltzmann ansatz}, namely the assumption that the pre-interaction states $\pr{v}$, $\pr{v_\ast}$ are statistically independent at the moment of the interaction.

The \textit{$s$-th statistical moment} of the distribution function $f$ is the quantity:
$$ M_s(t):=\int_\R v^sf(t,\,v)\,dv. $$
Multiplying~\eqref{eq:boltz.T} by $v^s$ and integrating over $\R$ yields the following equation for $M_s$:
$$ \frac{dM_s}{dt}=\int_\R\int_\R\left(\int_\R v^sT(v\,\vert\,\pr{v},\,\pr{v_\ast})\,dv\right)f(t,\,\pr{v})f(t,\,\pr{v_\ast})\,d\pr{v}\,d\pr{v_\ast}-M_s, $$
whence we deduce, in particular, that the \textit{mean} ($s=1$) and the \textit{energy} ($s=2$) of the system evolve according to
\begin{align*}
	\frac{dM_1}{dt} &= \int_\R\int_\R V_T(\pr{v},\,\pr{v_\ast})f(t,\,\pr{v})f(t,\,\pr{v_\ast})\,d\pr{v}\,d\pr{v_\ast}-M_1 \\[1mm]
	\frac{dM_2}{dt} &= \int_\R\int_\R E_T(\pr{v},\,\pr{v_\ast})f(t,\,\pr{v})f(t,\,\pr{v_\ast})\,d\pr{v}\,d\pr{v_\ast}-M_2,
\end{align*}
where
$$ V_T(\pr{v},\,\pr{v_\ast}):=\int_\R vT(v\,\vert\,\pr{v},\,\pr{v_\ast})\,dv, \qquad
	E_T(\pr{v},\,\pr{v_\ast}):=\int_\R v^2T(v\,\vert\,\pr{v},\,\pr{v_\ast})\,dv $$
denote the mean and the energy, respectively, of $T$ for a given pair $(\pr{v},\,\pr{v_\ast})\in\R\times\R$ of pre-interaction states.

\subsection{Collision-like models}
Modelling the interactions by providing such a detailed microscopic description of the system as imposed by the full specification of the transition probability density may actually not be straightforward. It may be doable if one resorts to relatively simple expressions of $T$, cf. e.g.~\cite{freguglia2017CMS,puppo2017KRM}. Nevertheless, as soon as one tries to make the microscopic model richer, cf. e.g.~\cite{bertotti2008MCM,dolfin2017KRM}, the expression of $T$ gets inevitably more complicated and often hardly amenable to deep analytical investigations, such as e.g. the identification of the asymptotic distributions (steady states) emerging from the interactions.

An alternative kinetic description~\cite{pareschi2013BOOK}, more closely inspired by the classical kinetic theory of gas dynamics, is based on microscopic ``\textit{collisions}'' expressed as
\begin{equation}
	v'=I(v,\,v_\ast)+D(v,\,v_\ast)\eta, \qquad v_\ast'=I(v_\ast,\,v)+D(v_\ast,\,v)\eta_\ast,
	\label{eq:binary_gen}
\end{equation}
where now $v,\,v_\ast\in\R$ denote the pre-collisional states of the interacting agents and $v',\,v_\ast'\in\R$ their post-collisional states. Here, $I:\R\times\R\to\R$ is a function modelling the \textit{deterministic part} of the collision, while $D:\R\times\R\to\R_+$ is a \textit{diffusion coefficient} expressing the intensity of the \textit{stochastic fluctuations} of the post-collisional states. Such stochastic fluctuations are modelled by the independent and identically distributed random variables $\eta,\,\eta_\ast\in\R$, which are taken with zero mean and unitary variance:
\begin{equation}
	\ave{\eta}=\ave{\eta_\ast}=0, \qquad \ave{\eta^2}=\ave{\eta_\ast^2}=1,
	\label{eq:eta.ave_var}
\end{equation}
the symbol $\ave{\cdot}$ denoting, here and henceforth, expectation with respect to the common law of $\eta$, $\eta_\ast$. Notice that the collisions~\eqref{eq:binary_gen} are \textit{symmetric}, in the sense that by switching $v$ and $v_\ast$ in the second rule one obtains the first rule. Therefore, in the following, it will be generally sufficient to refer only to the first rule in~\eqref{eq:binary_gen}.

If $f=f(t,\,v)$ is again the probability density function of the state $v$ at time $t$, the \textit{Boltzmann-type kinetic equation} ruling the evolution of $f$ under the microscopic model~\eqref{eq:binary_gen} is written as
\begin{equation}
	\partial_tf=\ave*{\int_\R\frac{1}{\pr{J}}f(t,\,\pr{v})f(t,\,\pr{v_\ast})\,dv_\ast}-f,
	\label{eq:boltz.strong}
\end{equation}
where the pre-collisional states $\pr{v}$, $\pr{v_\ast}$ have to be understood as functions of the post-collisional states $v$, $v_\ast$ according to the inverse of~\eqref{eq:binary_gen} and where $\pr{J}$ is the Jacobian of the transformation~\eqref{eq:binary_gen} as a function of the pre-collisional states. In order to avoid the necessity of invertible and smooth enough collision rules, it is customary to rewrite~\eqref{eq:boltz.strong} in weak form, namely to multiply it by a test function $\varphi:\R\to\R$, which in this context is called an \textit{observable quantity}, i.e. a quantity which may be expressed as a function of the microscopic state of the agents, and to integrate it on $\R$:
\begin{equation}
	\frac{d}{dt}\int_\R\varphi(v)f(t,\,v)\,dv=\int_\R\int_\R\ave{\varphi(v')-\varphi(v)}f(t,\,v)f(t,\,v_\ast)\,dv\,dv_\ast.
	\label{eq:boltz}
\end{equation}
In this form, the equation expresses the fact that the time variation of the expectation of $\varphi$ (left-hand side) is due to the mean variation of $\varphi$ in a binary collision (right-hand side). We will call~\eqref{eq:boltz.strong}, as well as its weak form~\eqref{eq:boltz}, a \textit{collisional kinetic equation}.

Choosing $\varphi(v)=v^s$, we get from~\eqref{eq:boltz} the evolution of the statistical moments of $f$. In particular, for $s=1$ we obtain
$$ \frac{dM_1}{dt}=\int_\R\int_\R I(v,\,v_\ast)f(t,\,v)f(t,\,v_\ast)\,dv\,dv_\ast-M_1, $$
while for $s=2$, considering that
\begin{align*}
	\ave{(v')^2-v^2} &= \ave{I^2(v,\,v_\ast)+2P(v,\,v_\ast)D(v,\,v_\ast)\eta+D^2(v,\,v_\ast)\eta^2-v^2} \\
	&= I^2(v,\,v_\ast)+D^2(v,\,v_\ast)-v^2,
\end{align*}
we discover
$$ \frac{dM_2}{dt}=\int_\R\int_\R\left(I^2(v,\,v_\ast)+D^2(v,\,v_\ast)\right)f(t,\,v)f(t,\,v_\ast)\,dv\,dv_\ast-M_2. $$
By comparing these results with the analogous ones found in Section~\ref{sect:jump_process}, we see that~\eqref{eq:boltz} provides the same evolution of the mean and of the energy as~\eqref{eq:boltz.T} if we choose
$$ I(v,\,v_\ast)=V_T(v,\,v_\ast), \qquad D(v,\,v_\ast)=\sqrt{E_T(v,\,v_\ast)-V_T^2(v,\,v_\ast)}=:D_T(v,\,v_\ast), $$
where $D_T(v,\,v_\ast)$ is the standard deviation of $T$ for a given pair $(v,\,v_\ast)\in\R\times\R$ of pre-collisional states. With these identifications, we may regard the collisional kinetic model~\eqref{eq:binary_gen}-\eqref{eq:boltz} as an alternative to the jump process kinetic model~\eqref{eq:boltz.T}, which reproduces correctly the evolution of the first two moments of the distribution function $f$. It is not superfluous to point out that these moments are typically involved in the hydrodynamic derivation of macroscopic equations from the kinetic description of the system.

Therefore, from now on we will consider the collisions
\begin{equation}
	v'=V_T(v,\,v_\ast)+D_T(v,\,v_\ast)\eta, \qquad v_\ast'=V_T(v_\ast,\,v)+D_T(v_\ast,\,v)\eta_\ast.
	\label{eq:binary}
\end{equation}

\section{Comparison of the kinetic descriptions}
\label{sect:comparisons}
Although models~\eqref{eq:boltz.T} and~\eqref{eq:boltz}-\eqref{eq:binary} account for the same evolution of the mean and of the energy, hence also of the variance, of $f$, they are in general not the same model for an arbitrary choice of the stochastic fluctuation $\eta$. In this section, starting from sufficiently simple motivating examples, which allow for detailed explicit computations, we show that, nonetheless, a proper choice of the law of $\eta$ actually makes the two models coincide.

\paragraph{Unconditional transition probability}
Assume that the transition probability density $T$ does not depend on the pre-interaction states $\pr{v}$, $\pr{v_\ast}$, thus $T=T(v)$. This is, for instance, the case considered in~\cite{chauviere2007NHM} to model cell-cell interactions in a biological context, $v$ being the cell velocity. Hence, the jump process model~\eqref{eq:boltz.T} simplifies to
$$ \partial_tf=T-f, $$
which, starting from an initial probability density $f_0$, admits for $t>0$ the solution
\begin{equation}
	f(t,\,v)=e^{-t}f_0(v)+\left(1-e^{-t}\right)T(v).
	\label{eq:f.a1}
\end{equation}

Concerning the collisional model~\eqref{eq:boltz}-\eqref{eq:binary}, we notice that in this case both $V_T$ and $D_T$ are constant, therefore the collision rule reduces to $v'=V_T+D_T\eta$. Let us consider $\varphi(v)=e^{-i\xi v}$ in~\eqref{eq:boltz}, where $i$ is the imaginary unit and $\xi\in\R$. Denoting by $\hat{f}(t,\,\xi):=\int_\R f(t,\,v)e^{-i\xi v}\,dv$ the Fourier transform of $f$ at time $t$, we obtain:
\begin{align*}
	\partial_t\hat{f} &= e^{-iV_T\xi}\ave*{e^{-iD_T\xi\eta}}-\hat{f}
\intertext{i.e., if $h=h(\eta):\R\to\R_+$ is the probability density of $\eta$,}
	&= e^{-iV_T\xi}\int_\R e^{-iD_T\xi\eta}h(\eta)\,d\eta-\hat{f} \\
	&= e^{-iV_T\xi}\hat{h}(D_T\xi)-\hat{f}.
\end{align*}
Going back to the physical variable $v$ by means of the inverse Fourier transform, we obtain the equation
$$ \partial_tf=\frac{1}{D_T}h\left(\frac{v-V_T}{D_T}\right)-f, $$
whose solution issuing from an initial probability density $f_0$ is
\begin{equation}
	f(t,\,v)=e^{-t}f_0(v)+\left(1-e^{-t}\right)\cdot\frac{1}{D_T}h\left(\frac{v-V_T}{D_T}\right).
	\label{eq:f.a2}
\end{equation}

Clearly,~\eqref{eq:f.a1} and~\eqref{eq:f.a2} are, in general, not the same kinetic distribution. Nevertheless, owing to the structure~\eqref{eq:binary} of the collision rules, they have the same mean and energy at every time $t\geq 0$ for every choice of the probability density $h$ complying with~\eqref{eq:eta.ave_var}. Moreover, if we let in particular
$$ h(\eta):=D_T T(D_T\eta+V_T) $$
then~\eqref{eq:f.a1} and~\eqref{eq:f.a2} become actually the same distribution. In practice, denoting by $V\sim T(\cdot)$ the random variable which represents the post-interaction state of an agent in the jump process model~\eqref{eq:boltz.T}, this amounts to choosing the stochastic fluctuation of the collisional model~\eqref{eq:binary} as
$$ \eta:=\frac{V-V_T}{D_T}, $$
namely as the standardisation of $V$, which indeed satisfies~\eqref{eq:eta.ave_var}.

Notice that the asymptotic distribution approached by either model for $t\to +\infty$, i.e. the equivalent of the \textit{Maxwellian} of the classical Boltzmann equation, is $T(v)$ in~\eqref{eq:f.a1} and $\frac{1}{D_T}h(\frac{v-V_T}{D_T})$ in~\eqref{eq:f.a2}. The convergence is in the $1$-Wasserstein metric and also in the $L^1$-norm if $f_0,\,T,\,g\in L^1(\R)$. Clearly, if $\eta$ is linked to $V$ as indicated above then the two Maxwellians coincide.

\paragraph{Run-and-tumble-like transition probability}
As a second example, we consider the case in which $T$ does not depend on the pre-interaction state $\pr{v_\ast}$ but may depend on $\pr{v}$, so that $T=T(v\,\vert\,\pr{v})$. This is, for instance, the formulation of the run-and-tumble model introduced in~\cite{alt1980JMB,stroock1974ZWVG} to describe the random motion of some species of bacteria, the variable $v$ representing the velocity of the bacteria. To be definite, let us consider e.g.
$$ T(v\,\vert\,\pr{v})=\frac{1}{2\sqrt{3}\lambda}\chi(v\in [\pr{v}-\sqrt{3}\lambda,\,\pr{v}+\sqrt{3}\lambda]), $$
where $\lambda>0$ is a parameter and $\chi$ denotes the characteristic function. Hence, the post-interaction velocity $v$ is uniformly distributed in a neighbourhood of amplitude $2\sqrt{3}\lambda$ of the pre-interaction velocity $\pr{v}$.

With $T$ given by this expression, let us multiply the jump process kinetic equation~\eqref{eq:boltz.T} by $e^{-i\xi v}$ and let us integrate it on $\R$. We obtain:
$$ \frac{d}{dt}\int_\R f(t,\,v)e^{-i\xi v}\,dv
	=\frac{1}{2\sqrt{3}\lambda}\int_\R\left(\int_{\pr{v}-\sqrt{3}\lambda}^{\pr{v}+\sqrt{3}\lambda}e^{-i\xi v}\,dv\right)
		f(t,\,\pr{v})\,d\pr{v}-\int_\R f(t,\,v)e^{-i\xi v}\,dv, $$
whence
$$ \partial_t\hat{f}=\left(\frac{\sin{(\sqrt{3}\lambda\xi)}}{\sqrt{3}\lambda\xi}-1\right)\hat{f}, $$
which, starting from a Fourier-transformable initial probability density $f_0$, yields
\begin{equation}
	\hat{f}(t,\,\xi)=\hat{f}_0(\xi)e^{\left(\frac{\sin{(\sqrt{3}\lambda\xi)}}{\sqrt{3}\lambda\xi}-1\right)t}.
	\label{eq:fhat.b1}
\end{equation}
Although it may be hard to obtain $f(t,\,v)$ explicitly by the inverse Fourier transform, this representation characterises $f$ univocally. Since $M_s(t)=i^s\partial_\xi^s\hat{f}(t,\,0)$, $s\in\N$, we compute in particular:
$$ M_1(t)=i\partial_\xi\hat{f}(t,\,0)=M_{1,0}, \qquad
	M_2(t)=-\partial^2_\xi\hat{f}(t,\,0)=M_{2,0}+\lambda^2 t, $$
where $M_{1,0}$, $M_{2,0}$ denote, respectively, the mean and the energy of the initial condition $f_0$. Therefore, we discover that the mean of the system is conserved, while the energy, hence also the variance, grows linearly in time at rate $\lambda^2$.

Concerning instead the collisional model~\eqref{eq:boltz}-\eqref{eq:binary}, we observe that, due to the structure of $T$, in this case we have $V_T(v,\,v_\ast)=V_T(v)$, $D_T(v,\,v_\ast)=D_T(v)$ with, in particular,
$$ V_T(v)=v, \qquad D_T(v)=\lambda, $$
thus the collision rule~\eqref{eq:binary} specialises as $v'=v+\lambda\eta$. From~\eqref{eq:boltz} with $\varphi(v)=e^{-i\xi v}$, we compute
$$ \partial_t\hat{f}=\ave*{e^{-i\xi\lambda\eta}-1}\hat{f}, $$
whence, denoting again by $h$ the distribution of $\eta$, we get
$$ \partial_t\hat{f}=\left(\hat{h}(\lambda\xi)-1\right)\hat{f}. $$
Starting from a Fourier-transformable initial probability density $f_0$, this yields
\begin{equation}
	\hat{f}(t,\,\xi)=\hat{f}_0(\xi)e^{\left(\hat{h}(\lambda\xi)-1\right)t}.
	\label{eq:fhat.b2}
\end{equation}
Also in this case we cannot obtain $f(t,\,v)$ explicitly by the inverse Fourier transform. Nevertheless, since this representation characterises $f$ univocally, we conclude that, in general, $f$ is not the same distribution as the one obtained with the jump process model. Yet, the theory developed in Section~\ref{sect:micro_int}, or alternatively a direct computation using the link between the derivatives of the Fourier transform and the moments of $f$ recalled before, reveals that, for every choice of $h$ (with $\hat{h}(0)=1$, $\hat{h}'(0)=0$, $\hat{h}''(0)=-1$ to be consistent with~\eqref{eq:eta.ave_var}), the trends of the first two statistical moments are exactly the same as those computed above.

The last point allows us to conclude that both models have invariably the same time-asymptotic trend. In fact, since in both cases the variance of $f$ grows unboundedly in time, the asymptotic distribution spreads on the whole real line independently of the initial condition. Hence, $f(t,\,v)\to 0$ for a.e. $v\in\R$ when $t\to +\infty$. Notice, however, that $f$ does not converge to $0$ in $L^1(\R)$, because $\Vert f(t,\,\cdot)\Vert_1=\hat{f}(t,\,0)=\hat{f}_0(0)=\Vert f_0\Vert_1=1$ for all $t\geq 0$.

Finally, we observe that~\eqref{eq:fhat.b1},~\eqref{eq:fhat.b2} are the Fourier transform of the same distribution for every $t\geq 0$ if
$$ h(\eta):=\frac{1}{2\sqrt{3}}\chi(\eta\in [-\sqrt{3},\,\sqrt{3}])=\lambda T(\lambda\eta+v\,\vert\, v), $$
for then it results $\hat{h}(\xi)=\frac{\sin(\sqrt{3}\xi)}{\sqrt{3}\xi}$. Once again, this corresponds to defining $\eta$ as the standardisation of $V$, i.e. $\eta:=\frac{V-v}{\lambda}$.

\paragraph{The general case}
If we admit that $\eta$ may depend on the pre-collisional states $v$, $v_\ast$, which thus parametrise its law, then we may generalise the previous results by setting
\begin{equation}
	\eta:=\frac{V-V_T(v,\,v_\ast)}{D_T(v,\,v_\ast)},
	\label{eq:eta_standard}
\end{equation}
which, considering that $V\sim T(\cdot\,\vert\,\pr{v},\,\pr{v_\ast})$, implies
$$ h(\eta\,\vert\, v,\,v_\ast):=D_T(v,\,v_\ast)T\bigl(D_T(v,\,v_\ast)\eta+V_T(v,\,v_\ast)\,\vert\, v,\,v_\ast\bigr). $$
We are implicitly assuming that $D_T(v,\,v_\ast)>0$ for all $v,\,v_\ast\in\R$, so as to avoid inessential technical difficulties with the standardisation~\eqref{eq:eta_standard}. Recalling~\eqref{eq:binary}, this produces
\begin{align*}
	\ave{\varphi(v')} &= \ave*{\varphi\bigl(V_T(v,\,v_\ast)+D_T(v,\,v_\ast)\eta\bigr)} \\
	&= \int_\R\varphi\bigl(V_T(v,\,v_\ast)+D_T(v,\,v_\ast)\eta\bigr)\cdot D_T(v,\,v_\ast)T\bigl(D_T(v,\,v_\ast)\eta+V_T(v,\,v_\ast)\,\vert\, v,\,v_\ast\bigr)\,d\eta,
\intertext{which, with the substitution $v':=V_T(v,\,v_\ast)+D_T(v,\,v_\ast)\eta$ for fixed $v,\,v_\ast\in\R$, yields}
	&= \int_\R\varphi(v')T(v'\,\vert\, v,\,v_\ast)\,dv'.
\end{align*}
Plugging this into~\eqref{eq:boltz} and renaming conveniently the dummy integration variables, we see that~\eqref{eq:boltz} reduces to the weak form of~\eqref{eq:boltz.T}, thereby confirming that the choice~\eqref{eq:eta_standard} makes the jump process model and the collisional model equivalent.

\begin{remark}
An important consequence of the equivalence between the kinetic models~\eqref{eq:boltz.T} and~\eqref{eq:boltz}-\eqref{eq:binary} with~\eqref{eq:eta_standard} is that the former may be simulated numerically by taking advantage of Monte Carlo algorithms typically used for the latter~\cite{bobylev2000PRE,pareschi2013BOOK}, though at the cost of sampling a stochastic fluctuation whose law changes with the pre-collisional states of the agents. See Appendix~\ref{app:MC}.
\label{rem:MC_T}
\end{remark}

\section{Quasi-invariant interactions}
\label{sect:quasi-invariant}
One of the main goals in the study of kinetic models is to characterise the stationary distributions arising asymptotically for $t\to +\infty$, for they depict the \textit{emergent behaviour} of the system of interacting agents.

As a matter of fact, the jump process model~\eqref{eq:boltz.T} hardly allows one to investigate in detail the trend to equilibrium. Some general results about the existence and uniqueness of equilibrium distributions are available, under possibly restrictive assumptions~\cite{freguglia2017CMS}. More detailed analytical investigations have been performed in the case of linear equations, i.e. when the transition probability $T$ does not depend on the pre-interaction state $\pr{v_\ast}$, cf.~\cite{hillen2000SIAP}. Nevertheless, typically the explicit expression of the Maxwellian $f_\infty$ can be inferred only in remarkable special cases, such as those dealt with in Section~\ref{sect:comparisons} and e.g. in~\cite{delitala2014KRM,puppo2017KRM}. In most situations, one needs to rely mainly on numerical simulations, which, on the other hand, may require very accurate numerical schemes in order to escape the trap of possible spurious equilibria, see e.g.~\cite{puppo2017CMS,puppo2017KRM}. For completeness, we report here, in the notation of this paper, the statement of the result proved in~\cite[Appendix A]{freguglia2017CMS} about the existence and uniqueness of equilibrium distributions for models of type~\eqref{eq:boltz.T}:
\begin{theorem}
Let the mapping $(\pr{v},\,\pr{v_\ast})\mapsto T(\cdot\,\vert\,\pr{v},\,\pr{v_\ast})$ be Lipschitz continuous with respect to the $1$-Wasserstein metric $W_1$ in the space of the probability measures on $\cI\subseteq\R$ (where $\cI$ may be either bounded or unbounded), i.e. let a constant $\Lip{(T)}>0$ exist such that
$$ W_1\bigl(T(\cdot\,\vert\,\pr{v},\,\pr{v_\ast}),\,T(\cdot\,\vert\,\pr{w},\,\pr{w_\ast})\bigr)\leq
	\Lip{(T)}\bigl(\abs{\pr{w}-\pr{v}}+\abs{\pr{w_\ast}-\pr{v_\ast}}\bigr),
		\qquad \forall\,\pr{v},\,\pr{v_\ast},\,\pr{w},\,\pr{w_\ast}\in\cI. $$
If $\Lip{(T)}<\frac{1}{2}$ then~\eqref{eq:boltz.T} admits a unique equilibrium distribution $f_\infty$, which is a probability measure on $\cI$ and which is also globally attractive, i.e.
$$ \lim_{t\to +\infty}W_1(f(t,\,\cdot),\,f_\infty)=0 $$
for every solution $f$ to~\eqref{eq:boltz.T}.
\end{theorem}

Conversely, the collisional model~\eqref{eq:boltz}-\eqref{eq:binary} offers several analytical tools, which often permit to recover explicitly accurate approximations of $f_\infty$ by means of suitable asymptotic procedures.  The basic idea of such procedures is to approximate the integro-differential equation~\eqref{eq:boltz} with partial differential equations, more amenable to analytical investigations, at least in some regimes of the parameters of the microscopic collisions. Indeed, identifying the equilibrium distributions directly from~\eqref{eq:boltz} by setting the right-hand side to zero remains a quite difficult, and often unsolvable, task.

A remarkable case in which this type of asymptotic analysis is successfully applied to~\eqref{eq:boltz} is that of the \textit{quasi-invariant interactions}. Let $\pr{V},\,V\in\R$ be the random variables representing the pre- and post-interaction states, respectively, of an agent and $\pr{V_\ast}\in\R$ the one representing the pre-interaction state of the other agent involved in the interaction. In the probabilistic description via the transition probabilities, we say that interactions are quasi-invariant if, given $0<\epsilon\ll 1$,
\begin{equation}
	\P(\abs{V-\pr{V}}>\epsilon\,\vert\,\pr{V},\,\pr{V_\ast})\leq\epsilon.
	\label{eq:quasi-inv_int}
\end{equation}
In other words, if the post-interaction state is, in probability, close to the pre-interaction state, so that the interactions produce a small transfer of microscopic state between the interacting agents. This concept was first introduced in the kinetic literature on multi-agent systems in~\cite{cordier2005JSP,toscani2006CMS} as a reminiscence of the \textit{grazing collisions} studied in the classical kinetic theory, see~\cite{villani1998ARMA}.

A class of transition probabilities $T$ satisfying~\eqref{eq:quasi-inv_int} is
\begin{equation}
	T(v\,\vert\,\pr{v},\,\pr{v_\ast})=(1-\epsilon)\delta(v-\pr{v})+\epsilon\cT(v\,\vert\,\pr{v},\,\pr{v_\ast}), \qquad 0<\epsilon\leq 1,
	\label{eq:T.eps}
\end{equation}
where $\cT(\cdot\,\vert\,\pr{v},\,\pr{v_\ast})$ is, for every $\pr{v},\,\pr{v_\ast}\in\R$, a probability density. Indeed, since $V\sim T(\cdot\,\vert\,\pr{v},\,\pr{v_\ast})$, it is straightforward to compute
$$ \P(\abs{V-\pr{V}}>\epsilon\,\vert\,\pr{V},\,\pr{V_\ast})=1-\P(\abs{V-\pr{V}}\leq\epsilon\,\vert\,\pr{V},\,\pr{V_\ast})
	=\epsilon\left(1-\int_{\pr{v}-\epsilon}^{\pr{v}+\epsilon}\cT(v\,\vert\,\pr{v},\,\pr{v_\ast})\,dv\right)\leq\epsilon, $$
thus the interactions described by~\eqref{eq:T.eps} are quasi-invariant. Roughly speaking,~\eqref{eq:T.eps} says that with probability $1-\epsilon$ an agent will not change his/her microscopic state after an interaction, whereas with probability $\epsilon$ s/he will change it according to the law encoded in $\cT$. In particular, it is easy to see that $T(\cdot\,\vert\,\pr{v},\,\pr{v_\ast})$ converges to $\delta(\cdot-\pr{v})$ in the $1$-Wasserstein metric as $\epsilon\to 0^+$, provided $\cT(\cdot\,\vert\,\pr{v},\,\pr{v_\ast})\in\mathscr{P}_1(\R)$ for every pair of pre-interaction states $(\pr{v},\,\pr{v_\ast})\in\R\times\R$ (see e.g.~\cite[Eq. (5.1.22)]{ambrosio2008BOOK} for the precise definition of the space $\mathscr{P}_1(\R)$).

Plugging~\eqref{eq:T.eps} into~\eqref{eq:boltz.T} yields
$$ \partial_tf=\epsilon\left(\int_\R\int_\R\cT(v\,\vert\,\pr{v},\,\pr{v_\ast})f(t,\,\pr{v})f(t,\,\pr{v_\ast})\,d\pr{v}\,d\pr{v_\ast}-f\right), $$
which indicates that the evolution of the jump process is now basically ruled by $\cT$ but slowed down by a factor $\epsilon$. This suggests that we introduce the new time scale $\tau:=\epsilon t$, which is much larger than the $t$-scale, hence more suited to show the large time behaviour of the system. Scaling consequently the distribution function as $g(\tau,\,v):=f(\frac{\tau}{\epsilon},\,v)$ and considering that $\partial_\tau g=\frac{1}{\epsilon}\partial_tf$, we see that $g$ satisfies the equation
\begin{equation}
	\partial_\tau g=\int_\R\int_\R\cT(v\,\vert\,\pr{v},\,\pr{v_\ast})g(\tau,\,\pr{v})g(\tau,\,\pr{v_\ast})\,d\pr{v}\,d\pr{v_\ast}-g,
	\label{eq:boltz.g-T}
\end{equation}
which is structurally identical to the very general equation~\eqref{eq:boltz.T}. Therefore, in spite of the quasi-invariant structure of the interactions, it is in principle not easier to extract from~\eqref{eq:boltz.g-T} any more detailed information about the asymptotic trends.

\subsection{Fokker-Planck asymptotic analysis}
Let us consider instead the collisional model~\eqref{eq:boltz}-\eqref{eq:binary} deduced from the transition probability~\eqref{eq:T.eps}. First of all, we compute:
\begin{align*}
	V_T(v,\,v_\ast) &= v+\epsilon(V_\cT(v,\,v_\ast)-v) \\
	E_T(v,\,v_\ast) &= V_T^2(v,\,v_\ast)+\epsilon(1-\epsilon){(V_\cT(v,\,v_\ast)-v)}^2+\epsilon D_\cT^2(v,\,v_\ast),
\intertext{where, obviously, $V_\cT(v,\,v_\ast)$ and $D_\cT(v,\,v_\ast)$ denote the mean and the standard deviation of the distribution $\cT$ for a fixed pair $(v,\,v_\ast)\in\R\times\R$ of pre-collisional states. From here, we also deduce:}
	D_T(v,\,v_\ast) &= \sqrt{\epsilon}\sqrt{(1-\epsilon){(V_\cT(v,\,v_\ast)-v)}^2+D_\cT^2(v,\,v_\ast)},
\end{align*}
thus, recalling~\eqref{eq:binary}, we are led to consider the collision rule
\begin{equation}
	v'=v+\epsilon(V_\cT(v,\,v_\ast)-v)+\sqrt{\epsilon}\sqrt{(1-\epsilon){(V_\cT(v,\,v_\ast)-v)}^2+D_\cT^2(v,\,v_\ast)}\eta.
	\label{eq:binary_quasi-inv}
\end{equation}
Notice that this collision is quasi-invariant, in the sense that if $\epsilon$ is small then $v'\approx v$.

In order to investigate the large time behaviour of~\eqref{eq:boltz} with the collision rule~\eqref{eq:binary_quasi-inv}, we scale again the time as $\tau:=\epsilon t$ and refer to the scaled distribution function $g(\tau,\,v)$ introduced above, which, in this case, satisfies the equation
\begin{equation}
	\frac{d}{d\tau}\int_\R\varphi(v)g(\tau,\,v)\,dv
		=\frac{1}{\epsilon}\int_\R\int_\R\ave{\varphi(v')-\varphi(v)}g(\tau,\,v)g(\tau,\,v_\ast)\,dv\,dv_\ast.
	\label{eq:boltz.g}
\end{equation}
We observe that, if $\varphi$ is sufficiently smooth, say $\varphi\in C^3(\R)$, when collisions are quasi-invariant we may expand:
\begin{align*}
	\frac{1}{\epsilon}\ave{\varphi(v')-\varphi(v)} &=
		\frac{1}{\epsilon}\ave*{\varphi'(v)(v'-v)+\frac{1}{2}\varphi''(v)(v'-v)^2+\frac{1}{6}\varphi'''(\bar{v})(v'-v)^3},
\intertext{where $\bar{v}\in (\min\{v,\,v'\},\,\max\{v,\,v'\})$, and further, using~\eqref{eq:binary_quasi-inv} and recalling~\eqref{eq:eta.ave_var}:}
	&= \varphi'(v)(V_\cT(v,\,v_\ast)-v)+\frac{1}{2}\varphi''(v)\left({(V_\cT(v,\,v_\ast)-v)}^2+D_\cT^2(v,\,v_\ast)\right)+o(\sqrt{\epsilon}),
\end{align*}
where $o(\sqrt{\epsilon})$ denotes, for every $v,\,v_\ast\in\R$, a remainder negligible with respect to $\sqrt{\epsilon}$, see~\cite{toscani2006CMS} for details. Therefore, the equation satisfied by $g$ takes the form
\begin{align*}
	\frac{d}{d\tau}\int_\R\varphi(v)g(\tau,\,v)\,dv &= \int_\R\int_\R\varphi'(v)\left(V_\cT(v,\,v_\ast)-v\right)g(\tau,\,v)g(\tau,\,v_\ast)\,dv\,dv_\ast \\
	&\phantom{=} +\frac{1}{2}\int_\R\int_\R\varphi''(v)\left({(V_\cT(v,\,v_\ast)-v)}^2
		+D_\cT^2(v,\,v_\ast)\right)g(\tau,\,v)g(\tau,\,v_\ast)\,dv\,dv_\ast \\
	&\phantom{=} +R_\varphi^\epsilon(g,\,g)(\tau),
\end{align*}
where $R_\varphi^\epsilon(g,\,g)$ is a remainder still negligible with respect to $\sqrt{\epsilon}$, i.e. $R_\varphi^\epsilon(g,\,g)=o(\sqrt{\epsilon})$, if, for instance, $V_\cT(v,\,v_\ast)$, $D_\cT(v,\,v_\ast)$ have a polynomial growth of some maximum degree $n\in\N$, $\eta$ has finite third order moment, i.e. $\ave{\abs{\eta}^3}<+\infty$, and $g$ has integrable moments of order $3\max\{1,\,n\}+\nu$ for some $\nu>0$. In fact, the detailed expression of $R_\varphi^\epsilon(g,\,g)$ is
\begin{align*}
	R_\varphi^\epsilon(g,\,g)(\tau)=\frac{1}{6}\int_\R\int_\R &\varphi'''(\bar{v})\Bigl[\epsilon^2{\left(V_\cT(v,\,v_\ast)-v\right)}^3 \\
	&+ 3\epsilon\left(V_\cT(v,\,v_\ast)-v\right)\left((1-\epsilon){(V_\cT(v,\,v_\ast)-v)}^2+D_\cT^2(v,\,v_\ast)\right) \\
	&+ \sqrt{\epsilon}\left((1-\epsilon){(V_\cT(v,\,v_\ast)-v)}^2+D_\cT^2(v,\,v_\ast)\right)^{3/2}\ave{\eta^3}\Bigr]g(\tau,\,v)g(\tau,\,v_\ast)\,dv\,dv_\ast.
\end{align*}
Under these assumptions, it results $R_\varphi^\epsilon(g,\,g)\to 0$ for $\epsilon\to 0^+$. In such a limit, which is called the \textit{quasi-invariant limit}~\cite{cordier2005JSP,toscani2006CMS}, we find therefore that $g$ solves
\begin{align}
	\begin{aligned}[b]
		\frac{d}{d\tau}\int_\R\varphi(v)g(\tau,\,v)\,dv &=
			\int_\R\int_\R\varphi'(v)\left(V_\cT(v,\,v_\ast)-v\right)g(\tau,\,v)g(\tau,\,v_\ast)\,dv\,dv_\ast \\
			&\phantom{=} +\frac{1}{2}\int_\R\int_\R\varphi''(v)\left(\left(V_\cT(v,\,v_\ast)-v\right)^2
				+D_\cT^2(v,\,v_\ast)\right)g(\tau,\,v)g(\tau,\,v_\ast)\,dv\,dv_\ast.
	\end{aligned}
	\label{eq:FP.weak}
\end{align}

Notice that, for every fixed $\tau>0$, if $\epsilon$ is small then $t=\frac{\tau}{\epsilon}$ is large, therefore the limit $\epsilon\to 0^+$ describes the large time behaviour of the collisional model~\eqref{eq:boltz}-\eqref{eq:binary_quasi-inv}. As a consequence, the solution $g$ to~\eqref{eq:FP.weak} is expected to match with $f$ for large times and, in particular, the asymptotic distribution function $g_\infty$ obtained for $\tau\to +\infty$ to match with the Maxwellian $f_\infty$ of the collisional model when in the latter $\epsilon$ is sufficiently small. The advantage of~\eqref{eq:FP.weak} is that, unlike~\eqref{eq:boltz}, it might not be an integro-differential equation. In fact, integrating by parts the right-hand side and invoking the arbitrariness of $\varphi$, we recognise that~\eqref{eq:FP.weak} is the weak form of the following \textit{Fokker-Planck equation:}
\begin{align}
	\begin{aligned}[b]
		\partial_\tau g &= \frac{1}{2}\partial^2_v\left\{\left[\int_\R\left(\left(V_\cT(v,\,v_\ast)-v\right)^2
			+D_\cT^2(v,\,v_\ast)\right)g(\tau,\,v_\ast)\,dv_\ast\right]g\right\} \\
		&\phantom{=} -\partial_v\left[\left(\int_\R V_\cT(v,\,v_\ast)g(\tau,\,v_\ast)\,dv_\ast-v\right)g\right],
	\end{aligned}
	\label{eq:FP}
\end{align}
provided $g$ satisfies suitable conditions for $v\to\pm\infty$, which ensure that the boundary terms appearing from the integration by parts vanish. For instance, one such sufficient condition is $g,\,\partial_vg\to 0$ quickly enough for $v\to\pm\infty$. Moreover, the specific expressions of $V_\cT$ and $D_\cT$ allow, in many cases, for an explicit computation of the remaining integrals in terms of moments of $g$, which finally reduces~\eqref{eq:FP} to a linear partial differential equation with possibly non-constant coefficients.

Before proceeding further, we notice that if we choose $\varphi(v)=v,\,v^2$ in~\eqref{eq:FP.weak} we get that the first and second moments of $g$ evolve as:
\begin{align*}
	\frac{dM_1}{d\tau} &= \int_\R\int_\R V_\cT(v,\,v_\ast)g(\tau,\,v)g(\tau,\,v_\ast)\,dv\,dv_\ast-M_1 \\
	\frac{dM_2}{d\tau} &= \int_\R\int_\R E_\cT(v,\,v_\ast)g(\tau,\,v)g(\tau,\,v_\ast)\,dv\,dv_\ast-M_2,
\end{align*}
where $E_\cT(v,\,v_\ast):=V_\cT^2(v,\,v_\ast)+D_\cT^2(v,\,v_\ast)$ is the energy of the distribution $\cT$ for a fixed pair $(v,\,v_\ast)\in\R\times\R$ of pre-collisional states. Therefore, we conclude that the Fokker-Planck equation reproduces correctly the large time evolution of the first two statistical moments of the jump process model~\eqref{eq:boltz.T}-\eqref{eq:T.eps}.

On the other hand, we observe that the asymptotic procedure leading to~\eqref{eq:FP.weak},~\eqref{eq:FP} is actually \textit{independent} of the specific law of $\eta$, the only important features being~\eqref{eq:eta.ave_var}. Therefore, we expect that the asymptotic distribution $g_\infty$ given by the Fokker-Planck equation~\eqref{eq:FP} may be, in general, only a possibly rough approximation of the Maxwellian of model~\eqref{eq:boltz.g-T}. Nevertheless, it may still be useful to catch qualitatively the big picture of the asymptotic trend, if the latter cannot be determined more accurately.

\begin{remark}
The quasi-invariant limit of the jump process model~\eqref{eq:boltz.T}-\eqref{eq:T.eps} is~\eqref{eq:boltz.g-T}, i.e. still a model in which particle interactions drive the dynamics of the system. Conversely, the quasi-invariant limit of the collisional model~\eqref{eq:boltz}-\eqref{eq:binary_quasi-inv} is~\eqref{eq:FP}, i.e. a model in which \textit{mean-field interactions} emerge as the long-run effect of the collisions. This explains why~\eqref{eq:FP} is independent of the specific law of $\eta$. On the other hand, from Section~\ref{sect:comparisons} we know that the law of $\eta$ is important in order to catch exactly the dynamics of the jump process model~\eqref{eq:boltz.T} by means of the collisional model~\eqref{eq:boltz}-\eqref{eq:binary}. Therefore, we understand why the mean-field model~\eqref{eq:FP} may be only a (possibly rough) approximation, however useful, of~\eqref{eq:boltz.g-T}.
\label{rem:approx}
\end{remark}

\subsection{Explicit determination of the Maxwellian}
\label{sect:Maxwellian}
We now review some meaningful examples, including the cases discussed in Section~\ref{sect:comparisons}, in the frame of the quasi-invariant interactions~\eqref{eq:T.eps},~\eqref{eq:binary_quasi-inv}, focusing specifically on the explicit determination of the asymptotic distribution.

\begin{figure}
\centering
\resizebox{0.495\textwidth}{!}{
\begin{tikzpicture}[
declare function={
	g(\x,\D)=2/(\pi*\D)*1/(1+((\x-\V)/\D)^2)^2);
} 
]
\pgfmathsetmacro{\V}{2}
\pgfmathsetmacro{\pi}{3.141592654}
\begin{axis}[
	axis on top,
	axis lines=middle,
	enlarge x limits=0.1,
	enlarge y limits=0.1,
	xtick={0,\V},
	xticklabels={0,$V_\cT$},
	ytick=\empty,
	xlabel={$v$}, x label style={at={(current axis.right of origin)},anchor=north},
	ylabel={$g_\infty(v)$}, y label style={at={(current axis.above origin)},anchor=east},
	xmin=-2,
	xmax=7,
	ymin=0,
	ymax=0.6,
	legend style={at={(1,1)},anchor=north east},
	legend entries={{$D_\cT^2=1$},{$D_\cT^2=4$},{$D_\cT^2=9$}}
]
\coordinate (O) at (0,0);
\addplot[orange,domain=-2.5:7.5,smooth,very thick,style=solid,mark=o,samples=50]{g(x,1)};
\addplot[cyan,domain=-2.5:7.5,smooth,very thick,style=solid,mark=square,samples=20]{g(x,2)};
\addplot[gray,domain=-2.5:7.5,smooth,very thick,style=solid,mark=*,samples=20]{g(x,3)};
\node at (O) [below left = 1pt of O] {$0$};
\end{axis}
\end{tikzpicture}
}
\caption{The distribution~\eqref{eq:ginf.1} with mean $V_\cT>0$ and increasing values of the variance $D_\cT^2$.}
\label{fig:ginf.1}
\end{figure}
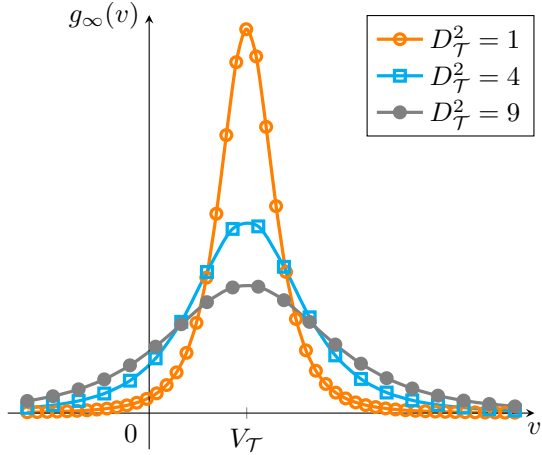

\paragraph{Unconditional dynamics}
If $\cT$ does not depend on the pre-interaction states $\pr{v},\,\pr{v_\ast}$, so that $V_\cT$ and $D_\cT$ are constant, the collision~\eqref{eq:binary_quasi-inv} reads
$$ v'=v+\epsilon(V_\cT-v)+\sqrt{\epsilon}\sqrt{(1-\epsilon)\left(V_\cT-v\right)^2+D_\cT^2}\eta $$
and the Fokker-Planck equation~\eqref{eq:FP} becomes
$$ \partial_\tau g=\frac{1}{2}\partial^2_v\left[\left({(V_\cT-v)}^2+D_\cT^2\right)g\right]-\partial_v\left((V_\cT-v)g\right). $$
The stationary distribution $g_\infty=g_\infty(v)$ solves
$$ \frac{1}{2}\partial_v\left[\left({(V_\cT-v)}^2+D_\cT^2\right)g_\infty\right]-(V_\cT-v)g_\infty=0, $$
whence we compute that the unique solution with unitary mass is
\begin{equation}
	g_\infty(v)=\frac{2}{\pi D_\cT\left[1+\left(\frac{v-V_\cT}{D_\cT}\right)^2\right]^2},
	\label{eq:ginf.1}
\end{equation}
see Figure~\ref{fig:ginf.1}. It can be checked that the mean and the variance of such a $g_\infty$ are the expected ones, i.e. $V_\cT$ and $D_\cT^2$, respectively. Moreover, since $g_\infty(v)\asymp v^{-4}$ for $v\to\pm\infty$, we observe that $g_\infty$ is a \textit{fat-tailed} distribution. Heuristically, this means that agents characterised asymptotically by a large value of $\abs{v}$ may be rare but not that improbable compared to what happens in classical physical phenomena, where the agent probability distribution typically decays exponentially fast at infinity.

From Section~\ref{sect:comparisons}, we know that the stationary solution to~\eqref{eq:boltz.g-T} is actually $g_\infty(v)=\cT(v)$. Clearly, this is different from the Maxwellian computed from the Fokker-Planck equation, apart from the very special case in which $\cT$ is chosen \textit{a priori} precisely as the distribution~\eqref{eq:ginf.1}. This confirms that, as already observed in Remark~\ref{rem:approx}, the Fokker-Planck equation~\eqref{eq:FP} depicts, in general, only an approximation of the large time behaviour of~\eqref{eq:boltz.g-T}, however with the right macroscopic trends (mean and energy).

\paragraph{Run-and-tumble-like dynamics}
Consider now a transition probability distribution $\cT$ independent of the pre-interaction state $\pr{v_\ast}$, like in the case of the run-and-tumble dynamics. Then $V_\cT=V_\cT(v)$ and $D_\cT=D_\cT(v)$, so that the Fokker-Planck equation~\eqref{eq:FP} becomes
\begin{equation}
	\partial_\tau g=\frac{1}{2}\partial^2_v\left[\left({(V_\cT(v)-v)}^2+D_\cT^2(v)\right)g\right]
		-\partial_v\left((V_\cT(v)-v)g\right)
	\label{eq:FP.no_vast}
\end{equation}
and, in particular, the stationary distribution $g_\infty$ solves
$$ \frac{1}{2}\partial_v\left[\left({(V_\cT(v)-v)}^2+D_\cT^2(v)\right)g_\infty\right]-(V_\cT(v)-v)g_\infty=0. $$
Setting $H(v):={(V_\cT(v)-v)}^2+D_\cT^2(v)$ for brevity, this equation produces the following general representation formula for $g_\infty$:
\begin{equation}
	g_\infty(v)=\frac{C}{H(v)}\exp{\left(2\int\frac{V_\cT(v)-v}{H(v)}\,dv\right)},
	\label{eq:ginf.2}
\end{equation}
where $C>0$ is a constant to be fixed so as to fulfil the normalisation condition $\int_\R g_\infty(v)\,dv=1$ and where the integral on the right-hand side denotes any antiderivative of $(V_\cT(v)-v)/H(v)$.

\begin{remark}
For a general $\cT=\cT(v\,\vert\,\pr{v})$, the jump process model~\eqref{eq:boltz.T}-\eqref{eq:T.eps} typically does not allow one to compute an asymptotic trend as detailed as~\eqref{eq:ginf.2}.
\end{remark}

Notice, however, that formula~\eqref{eq:ginf.2} may fail in some circumstances, for example if $V_\cT(v)-v$ is constant and $D_\cT(v)$ does not guarantee the integrability of $g_\infty$ on $\R$. This happens in the run-and-tumble-like case studied in Section~\ref{sect:comparisons}, i.e. with
$$ \cT(v\,\vert\,\pr{v})=\frac{1}{2\sqrt{3}\lambda}\chi(v\in[\pr{v}-\sqrt{3}\lambda,\,\pr{v}+\sqrt{3}\lambda]), $$
which, from~\eqref{eq:binary_quasi-inv}, induces the collision rule $v'=v+\sqrt{\epsilon}\lambda\eta$. The Fokker-Planck equation~\eqref{eq:FP} takes then the form
$$ \partial_\tau g=\frac{\lambda^2}{2}\partial^2_vg, $$
which, passing to the Fourier transform, is solved by $\hat{g}(\tau,\,\xi)=\hat{g}_0(\xi)e^{-\frac{\lambda^2}{2}\xi^2\tau}$, where $\hat{g}_0$ is the Fourier transform of the initial distribution $g_0$. In particular, for $g_0(v)=\delta(v)$, i.e. $\hat{g}_0(\xi)=1$, by the inverse Fourier transform we determine that $g$ is, at every time, a Gaussian:
$$ g(\tau,\,v)=\frac{1}{\sqrt{2\pi\lambda^2\tau}}e^{-\frac{v^2}{2\lambda^2\tau}} $$
with zero mean and variance equal to $\lambda^2\tau$. Then, of course, $g(\tau,\,v)\to 0$ for a.e. $v\in\R$ when $\tau\to +\infty$, so that the pointwise limit $g_\infty$ is actually not a probability density function. Notice, however, that $g$ does not converge to $g_\infty=0$ in $L^1(\R)$, because $\Vert g(\tau,\,\cdot)\Vert_1=1$ for all $\tau>0$. Recalling the results of Section~\ref{sect:comparisons}, we see that, in this case, the Fokker-Planck equation~\eqref{eq:FP} catches the qualitatively correct asymptotic trend of~\eqref{eq:boltz.g-T}.

\paragraph{Consensus/dissensus dynamics}
Finally, let us examine a case in which the transition probability $\cT$ features a full dependence on the pre-interaction states $\pr{v},\,\pr{v_\ast}$. To be definite, we take inspiration from problems of opinion formation, therefore we assume that the microscopic state $v$ represents the opinion of the agents about a certain issue. For convenience, we may think that $v>0$ denotes agreement while $v<0$ denotes disagreement. Unlike classical models of opinion formation, see e.g.~\cite{boudin2009M2NA,toscani2006CMS}, here we allow $v$ to span the whole real line rather than being confined to a bounded interval. In particular, we consider the following transition probability:
\begin{equation}
	\cT(v\,\vert\,\pr{v},\,\pr{v_\ast})=p\delta\Bigl(v-\bigl(\pr{v}+\gamma(\pr{v_\ast}-\pr{v})\bigr)\Bigr)
		+(1-p)\delta\Bigl(v-\bigl(\pr{v}-\gamma(\pr{v_\ast}-\pr{v})\bigr)\Bigr),
	\label{eq:T.opdyn}
\end{equation}
with $p,\,\gamma\in (0,\,1)$, which expresses the fact that:
\begin{itemize}
\item with probability $p$ there is consensus between the interacting agents, indeed $\pr{v}$ moves towards $\pr{v_\ast}$ because $v=\pr{v}+\gamma(\pr{v_\ast}-\pr{v})$;
\item with probability $1-p$ there is dissensus between the interacting agents, indeed $\pr{v}$ moves away from $\pr{v_\ast}$ because $v=\pr{v}-\gamma(\pr{v_\ast}-\pr{v})$.
\end{itemize}
Since
$$ V_\cT(v,\,v_\ast)=v+\gamma(2p-1)(v_\ast-v), \qquad D_\cT^2(v,\,v_\ast)=4\gamma^2p(1-p)(v_\ast-v)^2, $$
the collision rule~\eqref{eq:binary_quasi-inv} takes the form
\begin{equation}
	v'=v+\epsilon\gamma(2p-1)(v_\ast-v)+\sqrt{\epsilon}\gamma\sqrt{1-\epsilon(2p-1)^2}\abs{v_\ast-v}\eta.
	\label{eq:binary.opdyn}
\end{equation}

Let us investigate preliminarily the trend of the first two moments of the collisional model~\eqref{eq:boltz.g}. Since $\ave{v'-v}=\epsilon\gamma(2p-1)(v_\ast-v)$, choosing $\varphi(v)=v$ in~\eqref{eq:boltz.g} reveals
$$ \frac{dM_1}{d\tau}=\gamma(2p-1)\int_\R\int_\R(v_\ast-v)g(\tau,\,v)g(\tau,\,v_\ast)\,dv\,dv_\ast=0, $$
therefore the mean opinion is conserved in time. We will henceforth denote it by $m\in\R$.

Moreover, since $\ave{(v')^2-v^2}=\epsilon\gamma^2(v_\ast-v)^2+2\epsilon\gamma(2p-1)v(v_\ast-v)$, choosing $\varphi(v)=v^2$ in~\eqref{eq:boltz.g} yields
\begin{align*}
	\frac{dM_2}{d\tau} &= \gamma^2\int_\R\int_\R(v_\ast-v)^2g(\tau,\,v)g(\tau,\,v_\ast)\,dv\,dv_\ast \\
	&\phantom{=} +2\gamma(2p-1)\int_\R\int_\R v(v_\ast-v)g(\tau,\,v)g(\tau,\,v_\ast)\,dv\,dv_\ast \\
	&= 2\gamma(2p-1-\gamma)(m^2-M_2).
\end{align*}
From here, we see that the coefficient $2p-1-\gamma$ is of paramount importance to determine the asymptotic trend of $M_2$ and, consequently, that of the kinetic distribution function $g$.

If $2p-1-\gamma>0$, i.e. $0<\gamma<2p-1$ with $\frac{1}{2}<p<1$, then $M_2\to m^2$ when $\tau\to +\infty$. This implies that the Maxwellian $g_\infty$ has zero variance, thus it is necessarily $g_\infty(v)=\delta(v-m)$. In this situation, the agents reach a global consensus on the mean opinion $m$.

Conversely, if $2p-1-\gamma<0$, i.e. $\max\{0,\,2p-1\}<\gamma<1$, then $M_2\to +\infty$ when $\tau\to +\infty$ unless $M_2(0)=m^2$. This indicates that the previous Maxwellian becomes unstable: as soon as the initial energy of the system is different from $m^2$, or equivalently the initial variance of the system is greater than zero, the asymptotic distribution spreads on the whole real line, meaning that the dissensus prevails in the system.

\begin{figure}[!t]
\centering
\begin{tikzpicture}[
declare function={
	g(\v,\sigma)=2*\sigma^3/(\pi*(\v^2+\sigma^2))^2;
} 
]
\pgfmathsetmacro{\V}{2}
\pgfmathsetmacro{\pi}{3.141592654}
\begin{groupplot}[
	group style={group size= 2 by 1,horizontal sep=15mm},
	width=0.495\textwidth,
	legend cell align=left,
	legend pos=outer north east,
	legend style={draw=none}
]
\nextgroupplot[
	axis on top,
	axis lines=middle,
	enlarge x limits=0.1,
	enlarge y limits=0.1,
	xtick={0,0.5,1}, xticklabels={0,$\frac{1}{2}$,1},
	ytick={0,1},
	xlabel={$p$}, x label style={at={(current axis.right of origin)},anchor=north},
	ylabel={$\gamma$}, y label style={at={(current axis.above origin)},anchor=east},
	xmin=0,
	xmax=1.1,
	ymin=0,
	ymax=1.1
]
	\coordinate (O) at (0,0);
	\path[name path=axis] (axis cs:0,0) -- (axis cs:1,0);
	\addplot[black,domain=0.5:1,smooth,very thick,style={solid},name path=A]{2*x-1};
	\addplot[gray,style={dashed},name path=B] coordinates {(0,1) (1,1)};
	\addplot[gray,style={dashed}] coordinates {(1,0) (1,1)};
	\addplot[gray!30] fill between[of=axis and B,soft clip={domain=0:0.51}];
	\addplot[gray!30] fill between[of=A and B,soft clip={domain=0.501:1}];
	\addplot[cyan,opacity=0.3] fill between[of=axis and A,soft clip={domain=0.501:1}];
	\node at (O) [below left = 1pt of O] {$0$};
	\node at (axis cs: 0.3,0.65){\begin{minipage}{3cm}\centering Global \\ dissensus\end{minipage}};
	\node at (axis cs: 0.8,0.2){\begin{minipage}{3cm}\centering Global \\ consensus\end{minipage}};
	\addplot[draw=none,
			 postaction={decorate,
			 	decoration={text along path,text={Mild consensus},raise=1ex,text align={center}}
			 },domain=0.5:1]{2*x-1};
			 
\nextgroupplot[
	axis on top,
	axis lines=middle,
	enlarge x limits=0.1,
	enlarge y limits=0.1,
	xtick=\empty,
	ytick=\empty,
	xlabel={$v$}, x label style={at={(current axis.right of origin)},anchor=north},
	ylabel={$g_\infty(v)$}, y label style={at={(current axis.above origin)},anchor=east},
	xmin=-5,
	xmax=5,
	ymin=0,
	ymax=0.2,
	legend style={at={(1,1)},anchor=north east},
	legend entries={{$\sigma^2=1$},{$\sigma^2=4$},{$\sigma^2=9$}}
]
	\coordinate (O) at (0,0);
	\addplot[orange,smooth,very thick,style=solid,mark=o,samples=50]{g(x,1)};
	\addplot[cyan,smooth,very thick,style=solid,mark=square,samples=20]{g(x,2)};
	\addplot[gray,smooth,very thick,style=solid,mark=*,samples=20]{g(x,3)};
	\node at (O) [below left = 1pt of O] {$0$};
\end{groupplot}
\end{tikzpicture}
\caption{Left: the asymptotic configurations produced by~\eqref{eq:binary.opdyn} in the quasi-invariant regime. The black line is $\gamma=2p-1$. Right: the Maxwellian~\eqref{eq:ginf.3} with $m=0$ and increasing variance $\sigma^2$.}
\label{fig:ginf.3}
\end{figure}
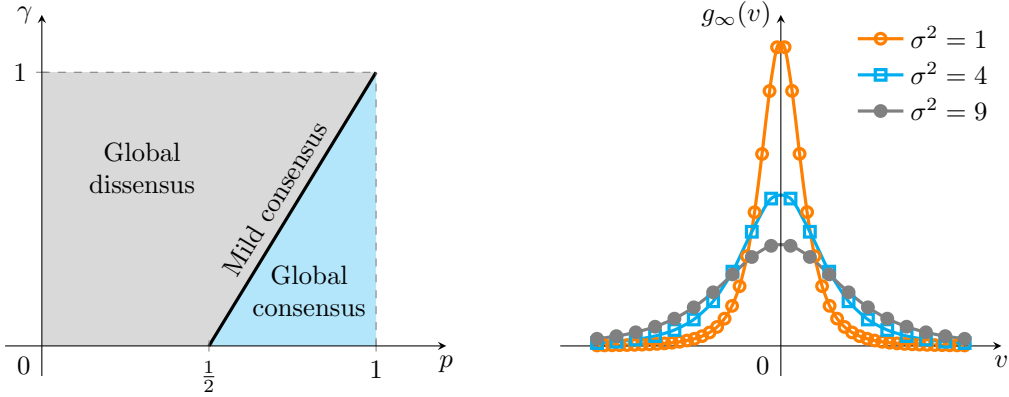

Finally, if $2p-1-\gamma=0$, i.e. $\gamma=2p-1$ with $\frac{1}{2}<p<1$, then also $M_2$ is conserved in time. In this case, let $\sigma^2:=M_2-m^2\in\R_+$ be the conserved variance of the distribution function $g$. We may investigate the asymptotic trend of the system taking advantage of the Fokker-Planck equation~\eqref{eq:FP}, which reads
$$ \partial_\tau g=(2p-1)^2\left[\frac{1}{2}\partial^2_v\Bigl(\bigl((v-m)^2+\sigma^2\bigr)g\Bigr)-\partial_v\bigl((m-v)g\bigr)\right]. $$
The unique solution with unitary mass to the stationary equation
$$ \frac{1}{2}\partial_v\Bigl(\bigl((v-m)^2+\sigma^2\bigr)g\Bigr)-(m-v)g=0 $$
identifies the Maxwellian:
\begin{equation}
	g_\infty(v)=\frac{2\sigma^3}{\pi\left[(v-m)^2+\sigma^2\right]^2},
	\label{eq:ginf.3}
\end{equation}
which, as expected, has mean $m$ and variance $\sigma^2$. In this case, a sort of mild consensus around the mean opinion $m$ emerges, which leaves the dispersion of the opinions unchanged. Notice that~\eqref{eq:ginf.3} is the same type of distribution as~\eqref{eq:ginf.1}, but its parameters have a substantially different meaning.

In Figure~\ref{fig:ginf.3}, we summarise the types of asymptotic configurations reached by the system depending on $\gamma$, $p$ and we plot some asymptotic distributions~\eqref{eq:ginf.3} with zero mean, indicating a neutral social opinion on average, and increasing variance.

It is worth pointing out that the qualitative diagram illustrated in the left panel of Figure~\ref{fig:ginf.3} applies actually also to the jump process model~\eqref{eq:boltz.g-T} with~\eqref{eq:T.opdyn}. In fact, since the trends of $M_1$, $M_2$ are the same, also the distribution function of this model either collapses to $\delta(v-m)$ (global consensus) or spreads on the whole $\R$ (global dissensus) or conserves both the mean and the energy (mild consensus) in correspondence of the same values of $\gamma$, $p$. Nonetheless, from~\eqref{eq:boltz.g-T}-\eqref{eq:T.opdyn} it is not as straightforward to find explicitly the expression of the Maxwellian in the less trivial case of mild consensus.

\begin{figure}[!t]
\centering
\includegraphics[width=0.7\textwidth]{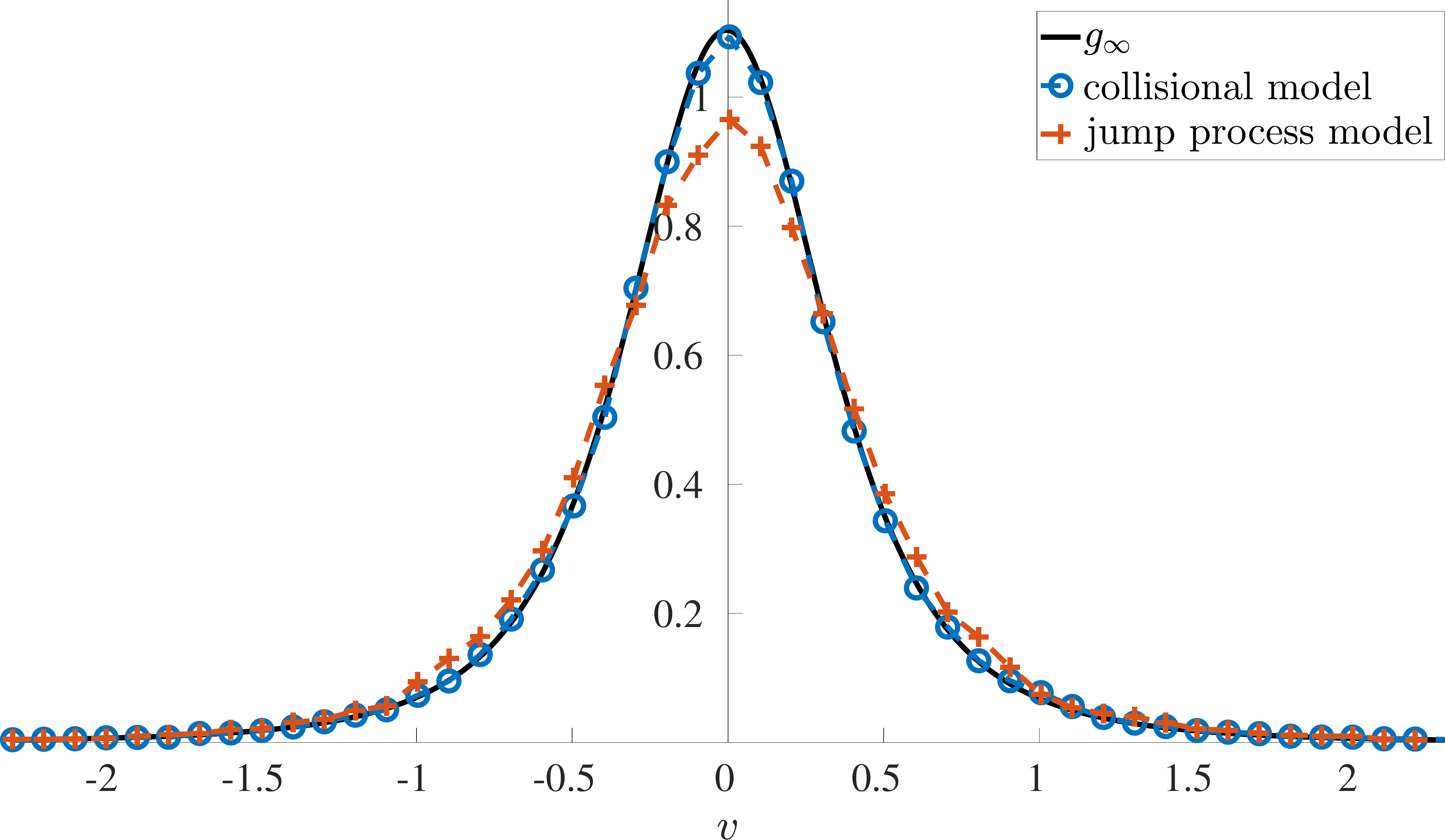}
\caption{Mild consensus: comparison among the large time solution to the jump process model~\eqref{eq:boltz.g-T}-\eqref{eq:T.opdyn} (cross markers), that to the collisional model~\eqref{eq:boltz.g}-\eqref{eq:binary.opdyn} (circular markers) and the analytical Maxwellian~\eqref{eq:ginf.3} (solid line). Parameters: $p=0.8$, $\gamma=2p-1=0.6$, $m=0$, $\sigma^2=0.3$, $\epsilon=10^{-3}$.}
\label{fig:ginf.3-num}
\end{figure}

In Figure~\ref{fig:ginf.3-num}, we show the numerical computation of the large time solution to the jump process model~\eqref{eq:boltz.g-T}-\eqref{eq:T.opdyn} in the case of mild consensus, obtained by means of a Monte Carlo algorithm (cf. Appendix~\ref{app:MC}). By comparing it with the numerical large time solution to the collisional model~\eqref{eq:boltz.g}-\eqref{eq:binary.opdyn}, obtained in turn with a Monte Carlo algorithm, and with the analytical Maxwellian~\eqref{eq:ginf.3}, we notice that, on one hand, the latter providesan excellent approximationof the large time collisional dynamics in the regime of quasi-invariant interactionsand, on the other hand, that the collisional description is indeed a valid tool to catch qualitatively the large time trend of the jump process model.

\subsection{Multidimensional extension}
For multidimensional problems, i.e. those in which the microscopic state of the agents is a vector variable $v\in\R^d$, $d>1$, the procedure leading to the Fokker-Planck equation may be repeated in a conceptually analogous way, though at the cost of more tedious computations.

Let us consider, for simplicity, only the case in which $\cT$ does not depend on the pre-interaction state $\pr{v_\ast}$. Denoting by $\V_\cT(\pr{v})\in\R^d$ the mean of $\cT$ for a given $\pr{v}$ and by
$$ \D_\cT(\pr{v}):=\int_{\R^d}\left(v-\V_\cT(\pr{v})\right)\otimes\left(v-\V_\cT(\pr{v})\right)T(v\,\vert\,\pr{v})\,dv\in\R^{d\times d} $$
its covariance matrix, the Fokker-Planck equation~\eqref{eq:FP.no_vast} generalises as
\begin{equation}
	\partial_\tau g=\frac{1}{2}\div\left\{\div\left[\Bigl((\V_\cT(v)-v)\otimes(\V_\cT(v)-v)
		+\D_\cT(v)\Bigr)g\right]\right\}-\div\left[\left(\V_\cT(v)-v\right)g\right],
	\label{eq:FP.multid}
\end{equation}
where $\div$ is the divergence operator with respect to the variable $v$. Clearly, this equation does not allow, in general, for the analytical computation of either $g$ or the Maxwellian $g_\infty$. Nevertheless, by means of suitable numerical schemes, see e.g.~\cite{loy2019PREPRINT,pareschi2017JSC}, one may compute accurate approximations of the steady states of~\eqref{eq:FP.multid} with much higher precision and efficiency than by tackling numerically the original collisional equation~\eqref{eq:boltz.g}.

\section{Domains with restrictions}
\label{sect:bounded_domains}
So far, we have considered only the case in which the microscopic state $v$ may span the entire real line. In many problems, however, the physically meaningful values of $v$ are restricted to a proper subset $\cI\subsetneq\R$, which may have, for instance, one of these prototypical forms:
\begin{equation}
	\cI=[-1,\,1], \qquad \cI=[0,\,1], \qquad \cI=\R_+.
	\label{eq:I}
\end{equation}
In order for both the pre- and the post-interaction states to belong to $\cI$, it is necessary that the transition probability density $T$ be supported in $\cI$ for every pair of pre-interaction states. We write:
$$ \supp{T(\cdot\,\vert\,\pr{v},\,\pr{v_\ast})}\subseteq \cI \quad \forall\,\pr{v},\,\pr{v_\ast}\in\cI, $$
so that $\int_\cI T(v\,\vert\,\pr{v},\,\pr{v_\ast})\,dv=1$ for all $\pr{v},\,\pr{v_\ast}\in\cI$. Notice that, by definition of mean, this implies
$$ V_T(\pr{v},\,\pr{v_\ast})\in\cI \quad \forall\,\pr{v},\,\pr{v_\ast}\in\cI. $$

As far as the description with the collision rules~\eqref{eq:binary} is concerned, if $\eta$, $\eta_\ast$ are chosen like in~\eqref{eq:eta_standard} then it results $v',\,v_\ast'\in\cI$ for all $v,\,v_\ast\in \cI$ by construction. Thus, the choice~\eqref{eq:eta_standard} guarantees straightforwardly the physical consistency of the collisional model also in subsets of $\R$.

If, conversely, $\eta$, $\eta_\ast$ are not chosen like in~\eqref{eq:eta_standard}, for instance because the collisional model~\eqref{eq:boltz}-\eqref{eq:binary} is of interest by itself, independently of the equivalence with the jump process model~\eqref{eq:boltz.T}, then, in general, $v'$, $v_\ast'$ may not belong to $\cI$, even if $v$, $v_\ast$ do. In order to deal with this issue, an option is to correct the Boltzmann-type equation~\eqref{eq:boltz} as follows, see~\cite{pareschi2013BOOK}:
\begin{equation}
	\frac{d}{dt}\int_\cI\varphi(v)f(t,\,v)\,dv
		=\frac{1}{2}\int_\cI\int_\cI B(v,\,v_\ast)\ave*{\varphi(v')+\varphi(v_\ast')-\varphi(v)-\varphi(v_\ast)}f(t,\,v)f(t,\,v_\ast)\,dv\,dv_\ast,
	\label{eq:boltz.beta}
\end{equation}
where $B:\cI\times \cI\to\R_+$ is a \textit{collision kernel} of the form
\begin{equation}
	B(v,\,v_\ast)=\chi(v'\in\cI)\chi(v_\ast'\in\cI).
	\label{eq:B}
\end{equation}
Notice that if the collisions~\eqref{eq:binary} actually ensure that $v',\,v_\ast'\in\cI$ then $B\equiv 1$ and~\eqref{eq:boltz.beta} reduces to~\eqref{eq:boltz} thanks to the symmetry of the rules~\eqref{eq:binary}. Conversely, if, for certain values of $v,\,v_\ast\in \cI$, the collisions~\eqref{eq:binary} produce either $v'\not\in \cI$ or $v_\ast'\not\in \cI$ then the kernel~\eqref{eq:B} has the effect of discarding those collisions from the microscopic dynamics of the system.

Unfortunately, because of $B$, the Boltzmann-type equation~\eqref{eq:boltz.beta} is much harder than~\eqref{eq:boltz} to treat, as far as the study of both the statistical moments and the asymptotic trend of $f$ are concerned. Actually, in some cases it may still be possible to derive Fokker-Planck equations in the quasi-invariant limit, see~\cite{cordier2005JSP,tosin2019MCRF_preprint}. However, the whole analysis is greatly simplified if one may find conditions on the collision rules~\eqref{eq:binary} ensuring $v',\,v_\ast'\in\cI$ for all $v,\,v_\ast\in \cI$, so as to get rid of the kernel $B$ and recover at once all the results valid in the case $\cI=\R$.

With this idea in mind, let us consider the first case in~\eqref{eq:I}, i.e. $\cI=[-1,\,1]$. This is the prototype of a symmetric interval about $v=0$ for problems in which both the positive and the negative values of the microscopic state are physically meaningful, provided they are bounded. An example is given by opinion dynamics models, where $v\in [-1,\,1]$ represents the opinion of the agents~\cite{boudin2009M2NA,toscani2006CMS}. We claim that it is actually possible to guarantee $v',\,v_\ast'\in [-1,\,1]$ for all $v,\,v_\ast\in [-1,\,1]$ by simply considering \textit{compactly supported} random variables $\eta$, $\eta_\ast$ in~\eqref{eq:binary}. First, we observe that, since $V_T(v,\,v_\ast)\in [-1,\,1]$ and
$$ E_T(v,\,v_\ast)=\int_{-1}^1 (v')^2T(v'\,\vert\, v,\,v_\ast)\,dv'\leq 1, $$
we have
$$ D_T(v,\,v_\ast)=\sqrt{E_T(v,\,v_\ast)-V_T^2(v,\,v_\ast)}\leq\sqrt{1-V_T^2(v,\,v_\ast)}\leq 1. $$
Therefore, the condition $v'\in [-1,\,1]$, or equivalently $\abs{v'}\leq 1$, is certainly satisfied if $\abs{V_T(v,\,v_\ast)}+D_T(v,\,v_\ast)\abs{\eta}\leq 1$ and even more so if $\abs{V_T(v,\,v_\ast)}+\abs{\eta}\leq 1$. If we now assume that there exists a constant $c\in (0,\,1)$ such that
$$ \abs{V_T(v,\,v_\ast)}\leq c \quad \forall\,v,\,v_\ast\in [-1,\,1], $$
we see that the restriction $\abs{\eta}\leq 1-c$ is sufficient to guarantee that $v'$ complies with the stated bounds. Analogously, if $\abs{\eta_\ast}\leq 1-c$ then $v_\ast'\in [-1,\,1]$. Finally, we conclude that if the random variables $\eta$, $\eta_\ast$ are such that
$$ \supp{\eta},\,\supp{\eta_\ast}\subseteq [-(1-c),\,1-c], \quad 0<c<1, $$
then the post-collisional states $v'$, $v_\ast'$ never violate the prescribed bounds. We observe that it is quite important that the obtained restriction on the supports of $\eta$, $\eta_\ast$ includes both positive and negative values, for otherwise it would be impossible for $\eta$, $\eta_\ast$ to comply with~\eqref{eq:eta.ave_var}.

Let us now consider the second case in~\eqref{eq:I}, i.e. $\cI=[0,\,1]$. This is the prototype of a microscopic state space for problems in which the microscopic state is non-negative and bounded. Examples include the speed of the vehicles or the safety of the drivers in vehicular traffic models~\cite{freguglia2017CMS,tosin2019MMS}; the social opinions in opinion dynamics models~\cite{bertotti2008MCM}; the estimated values of some traded goods in market dynamics models~\cite{delitala2014KRM}. In this case,
$$ E_T(v,\,v_\ast)=\int_0^1(v')^2T(v'\,\vert\, v,\,v_\ast)\,dv'\leq\int_0^1 v'T(v'\,\vert\, v,\,v_\ast)\,dv'=V_T(v,\,v_\ast), $$
hence
$$ D_T(v,\,v_\ast)=\sqrt{E_T(v,\,v_\ast)-V_T^2(v,\,v_\ast)}\leq\sqrt{V_T(v,\,v_\ast)(1-V_T(v,\,v_\ast))}\leq\frac{1}{2} $$
because $V_T(v,\,v_\ast)\in [0,\,1]$. Let us assume that there exists a constant $c\in (0,\,1)$ such that
\begin{equation}
	c\leq V_T(v,\,v_\ast)\leq 1-c \quad \forall\,v,\,v_\ast\in [0,\,1].
	\label{eq:VT_[0,1]}
\end{equation}
We observe that the condition $v'\geq 0$ is equivalent to $\eta\geq -\frac{V_T(v,\,v_\ast)}{D_T(v,\,v_\ast)}$, which, considering that $\frac{V_T(v,\,v_\ast)}{D_T(v,\,v_\ast)}\geq 2c$, is further enforced by imposing $\eta\geq -2c$. Analogously, the condition $v'\leq 1$ is equivalent to $\eta\leq\frac{1-V_T(v,\,v_\ast)}{D_T(v,\,v_\ast)}$, which, since $1-V_T(v,\,v_\ast)\geq c$ and $\frac{1}{D_T(v,\,v_\ast)}\geq 2$, is further enforced by imposing $\eta\leq 2c$. The same conclusions hold also for the random variable $\eta_\ast$. Summarising, if $\eta,\,\eta_\ast$ are chosen in such a way that
$$ \supp{\eta},\,\supp{\eta_\ast}\subseteq [-2c,\,2c], \quad 0<c<1, $$
then $v',\,v_\ast'\in [0,\,1]$ for all $v,\,v_\ast\in [0,\,1]$. Again, we notice that the restriction of $\eta$, $\eta_\ast$ to a bounded interval including both positive and negative values is essential for consistency with~\eqref{eq:eta.ave_var}.

Finally, let us consider the third case in~\eqref{eq:I}, i.e. $\cI=\R_+$. This is the prototype of the domain of a non-negative but possibly unbounded microscopic state, which is found e.g. in wealth distribution problems~\cite{cordier2005JSP}, where $v$ represents the wealth of the agents; in vehicular traffic problems~\cite{prigogine1960OR,prigogine1971BOOK}, where $v$ represents the speed of the vehicles; in models of human behaviour~\cite{gualandi2019M3AS}, where $v$ may represent e.g. the service time of the operators of a call centre~\cite{gualandi2018M3AS}, the number of inhabitants of a city~\cite{gualandi2019PHYSA}, the popularity of a product posted on a social network~\cite{toscani2018PRE} or various other social determinants. In this case, if we assume that there exist two constants $c_1>0$ and $0\leq c_2<+\infty$ such that
$$ V_T(v,\,v_\ast)\geq c_1, \quad D_T(v,\,v_\ast)\leq c_2 \qquad \forall\,v,\,v_\ast\in\R_+ $$
then, upon defining $c:=\frac{c_1}{c_2}>0$, we see that the condition $v'\geq 0$, i.e. $\eta\geq -\frac{V_T(v,\,v_\ast)}{D_T(v,\,v_\ast)}$, is certainly satisfied if one takes $\eta\geq -c$. Therefore, with
$$ \supp{\eta},\,\supp{\eta_\ast}\subseteq [-c,\,+\infty), \quad c>0, $$
we guarantee that $v',\,v_\ast'\in\R_+$ for all $v,\,v_\ast\in\R_+$ along with the consistency with~\eqref{eq:eta.ave_var}.

The results just presented are sufficient conditions valid for very general expressions of $V_T$, $D_T$. As such, they rely on some assumptions, such as the boundedness of $V_T$ away from the boundaries of $\cI$ or that of $D_T$, which may not be met in particular cases. Nevertheless, one may often take advantage of the specific expressions of $V_T$, $D_T$ to still find suitable bounds on $\eta$, $\eta_\ast$ which make the collisional model~\eqref{eq:binary} physically consistent.

\paragraph{Random onset of Alzheimer's disease in the brain}
For instance, in the model for the onset and progression of Alzheimer's disease presented in~\cite{bertsch2017MMB,bertsch2017JPA,bertsch2018SIMA}, one of the aspects taken into account is the random damage of the neurons due to either external or genetic factors. This aspect is modelled by assuming that the degree of malfunctioning $v\in [0,\,1]$ of a neuron (where $v=0$ stands for a healthy neuron and $v=1$ for a dead neuron) may randomly jump to a higher value according to the transition probability density
\begin{equation}
	T(v\,\vert\,\pr{v})=\frac{2}{1-\pr{v}}\chi\left(v\in\left[\pr{v},\,\frac{1+\pr{v}}{2}\right]\right).
	\label{eq:alzheimer_T}
\end{equation}
Notice that $\supp{T(\cdot\,\vert\,\pr{v})}\subseteq [0,\,1]$ for all $\pr{v}\in [0,\,1]$. As it can be easily computed, it results
$$ V_T(\pr{v})=\frac{3\pr{v}+1}{4}, \qquad D_T(\pr{v})=\frac{1-\pr{v}}{4\sqrt{3}}, $$
therefore, from~\eqref{eq:binary}, the corresponding collision rule is
\begin{equation}
	v'=\frac{3v+1}{4}+\frac{1-v}{4\sqrt{3}}\eta.
	\label{eq:alzheimer_collision}
\end{equation}
Since $V_T(1)=1$, it is clear that~\eqref{eq:VT_[0,1]} cannot be satisfied. Nevertheless, by rewriting the collision rule as
$$ v'=\frac{1}{4}\left(3-\frac{\eta}{\sqrt{3}	}\right)v+\frac{1}{4}\left(1+\frac{\eta}{\sqrt{3}}\right), $$
we see that we can guarantee $v'\in [0,\,1]$ by requiring
\begin{equation}
	0\leq\frac{1}{4}\left(1+\frac{\eta}{\sqrt{3}}\right)\leq 1 \quad \text{i.e.} \quad -\sqrt{3}\leq\eta\leq 3\sqrt{3}.
	\label{eq:alzheimer_eta}
\end{equation}
In fact, since $v'$ is a linear-affine function of $v$, it is necessary and sufficient that $v'\in [0,\,1]$ for $v=0,\,1$ in order for $v'\in [0,\,1]$ for all $v\in [0,\,1]$. The bounds obtained on $\eta$ are still consistent with~\eqref{eq:eta.ave_var}.

With these restrictions on $\eta$, the evolution of the collisional model is described by a Boltzmann-type equation of the form~\eqref{eq:boltz} but on $\cI=[0,\,1]$ rather than on $\R$:
$$ \frac{d}{dt}\int_0^1\varphi(v)f(t,\,v)\,dv=\int_0^1\ave{\varphi(v')-\varphi(v)}f(t,\,v)\,dv, $$
where the right-hand side is actually linear in $f$, because $v'$ does not depend on the pre-collisional state $v_\ast$. From here, taking $\varphi(v)=v$, we obtain that the average degree of malfunctioning of the neurons evolves according to the equation
$$ \frac{dM_1}{dt}=\frac{1}{4}\left(1-M_1\right), $$
which gives
\begin{equation}
	M_1(t)=\left(M_{1,0}-1\right)e^{-\frac{t}{4}}+1
	\label{eq:alzheimer_M1}
\end{equation}
with $M_{1,0}:=M_1(0)$. Analogously, taking $\varphi(v)=v^2$, we obtain that the energy of the system is ruled by the equation
$$ \frac{dM_2}{dt}=-\frac{5}{12}M_2+\frac{1}{3}M_1+\frac{1}{12}, $$
whose solution is
\begin{equation}
	M_2(t)=\left[M_{2,0}+2\left(M_{1,0}-1\right)\left(e^\frac{t}{6}-1\right)-1\right]e^{-\frac{5}{12}t}+1
	\label{eq:alzheimer_M2}
\end{equation}
with $M_{2,0}:=M_2(0)$.

\begin{figure}[!t]
\centering
\includegraphics[width=0.9\textwidth]{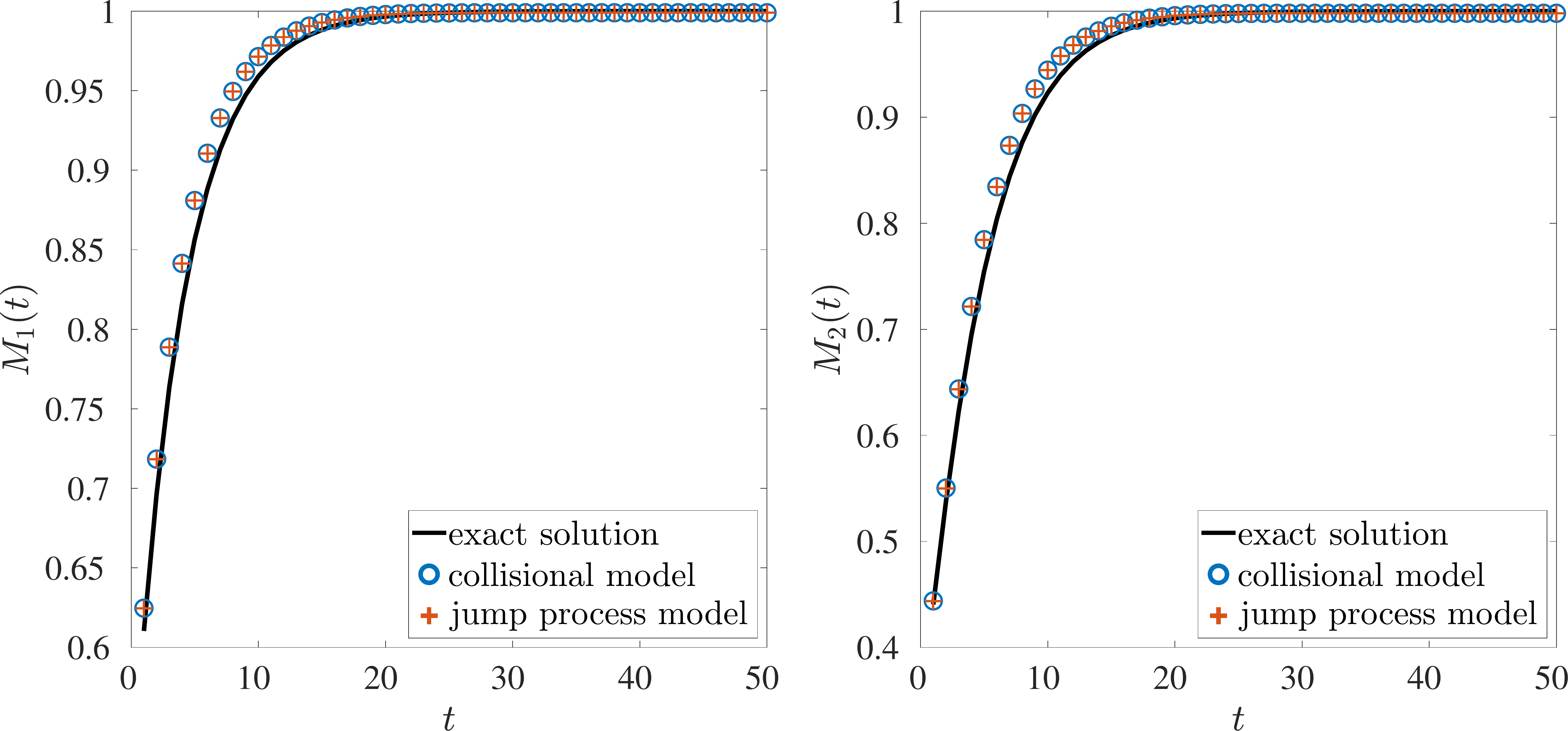}
\caption{Random onset of Alzheimer's disease: time evolution of $M_1$ (left), $M_2$ (right) computed numerically (circular and cross markers) and analytically from~\eqref{eq:alzheimer_M1},~\eqref{eq:alzheimer_M2} (solid line).}
\label{fig:alzheimer_M1M2}
\end{figure}

In Figure~\ref{fig:alzheimer_M1M2}, we show the time evolutions of $M_1$, $M_2$ computed numerically via a Monte Carlo algorithm (cf. Appendix~\ref{app:MC}) for both the collisional model and the jump process model~\eqref{eq:boltz.T} on $\cI=[0,\,1]$, i.e.
\begin{equation}
	\partial_tf=\int_0^1 T(v\,\vert\,\pr{v})f(t,\,\pr{v})\,d\pr{v}-f.
	\label{eq:alzheimer_jpke}
\end{equation}
Furthermore, we compare them with the analytical expressions~\eqref{eq:alzheimer_M1},~\eqref{eq:alzheimer_M2}. As expected from the general theory, we see that the two models predict the same instantaneous evolutionof these two statistical moments.

Notice that $M_1,\,M_2\to 1$ when $t\to +\infty$. Therefore, the variance $M_2-M_1^2$ of the equilibrium distribution is zero, whence we deduce that the Maxwellian is necessarily a Dirac delta centred in the asymptotic mean:
\begin{equation}
	f_\infty(v)=\delta(v-1)
	\label{eq:finf.delta1}
\end{equation}
for both models. This describes the fact that successive random damages of the neurons lead inevitably to a full damage of the brain, no matter what the initial condition is. Remarkably, in this case the specific choice of the distribution of $\eta$ in the collisional model is unimportant for the asymptotic equivalence of the two kinetic models.

\begin{figure}[!t]
\centering
\includegraphics[width=\textwidth]{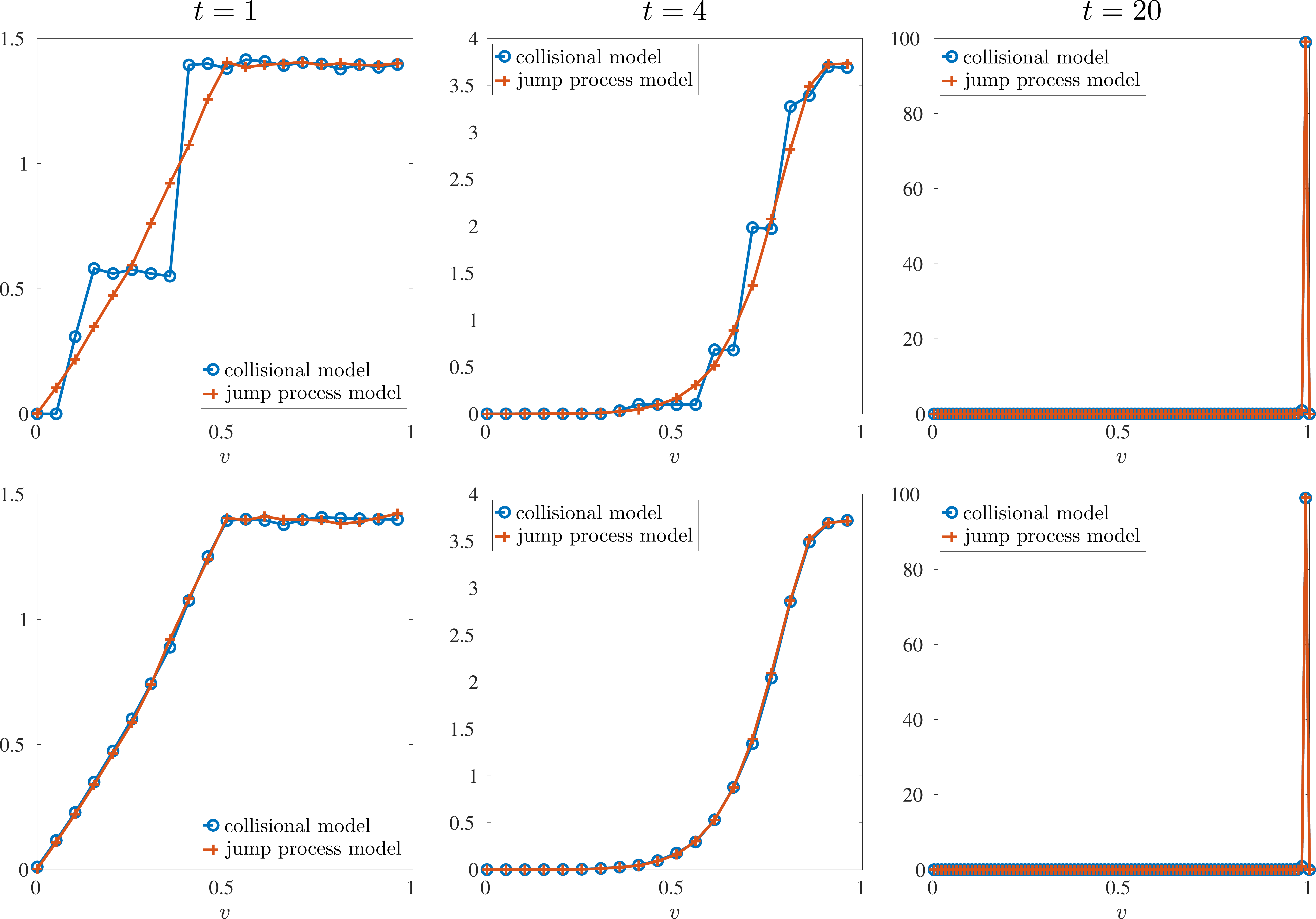}
\caption{Random onset of Alzheimer's disease: transient solutions, at the computational times $t=1,\,4,\,20$, of the jump process model (cross markers) and of the collisional model (circular markers). Top row: discrete $\eta$. Bottom row: uniformly distributed $\eta$.}
\label{fig:alzheimer_evo}
\end{figure}

Conversely, the choice of $\eta$ may affect the similarity of the transient solutions of the two kinetic models, as we show in Figure~\ref{fig:alzheimer_evo}. Starting from the initial condition $f_0(v)=\delta(v)$, which models a fully healthy brain, we consider, in the top row, a discrete stochastic fluctuation $\eta\in\{-1,\,1\}$ with law $\P(\eta=\pm 1)=\frac{1}{2}$. In the bottom row, we consider instead a uniform stochastic fluctuation $\eta\sim\cU([-\sqrt{3},\,\sqrt{3}])$, which corresponds to the standardisation~\eqref{eq:eta_standard} of the transition probability~\eqref{eq:alzheimer_T}. The displayed numerical solutions have been computed by means of a Monte Carlo algorithm (cf. Appendix~\ref{app:MC}). Both choices of $\eta$ are consistent with~\eqref{eq:alzheimer_eta} and, as expected, produce an evolution towards the Maxwellian~\eqref{eq:finf.delta1}. Nevertheless, we observe that the transient evolution of the collisional model produced by the discrete $\eta$ is, at the beginning, quite different from that of the jump process model. On the contrary, the transient evolution produced by the uniformly distributed $\eta$ is immediately the same (at least within the tolerance of a stochastic numerical method) as that of the jump process model.

\section{Conclusions}
\label{sect:conclusions}
In this paper, we have investigated two different classes of stochastic microscopic models of interacting agents and their corresponding kinetic descriptions. On one hand, we have considered microscopic jump process models, in which the transition from a pre-interaction state $\pr{v}$ to a post-interaction state $v$ of an agent interacting with another agent with state $\pr{v_\ast}$ is modelled by means of a conditional transition probability density $T=T(v\,\vert\,\pr{v},\,\pr{v_\ast})$. On the other hand, we have considered collision-like models, in which the post-interaction state $v'$ depends on the pre-interaction states $v$, $v_\ast$ of the interacting agents via a collision rule of the form $v'=I(v,\,v_\ast)+D(v,\,v_\ast)\eta$, where $I$ represents the deterministic part of the collision while $\eta$ is a stochastic fluctuation and $D$ a diffusion coefficient. The first type of models is especially used in biological applications, such as e.g. cell motion, see~\cite{alt1980JMB,chauviere2007NHM,hillen2006JMB,hillen2000SIAP}. The second type of models, which takes inspiration from the classical kinetic theory of gases, is instead typically used in socio-economic applications, see~\cite{pareschi2013BOOK} as a representative example, although in some cases socio-economic systems have been described also by means of jump process models, see e.g.~\cite{bertotti2008MCM,delitala2014KRM,dolfin2017KRM,puppo2017KRM}.

We have established a parallelism between the corresponding aggregate kinetic descriptions, whereby we have provided precise conditions for the equivalence of the two classes of models. First, we have shown that, by suitably linking the functions $I$, $D$ to the first and second order moments of $T$, the two kinetic descriptions account for the same evolution of the mean and of the energy of the system. Moreover, if, in the collisional model, one considers a suitably defined agent-dependent stochastic fluctuation $\eta$, namely one which changes from pair to pair of interacting agents on the basis of their pre-interaction states, then the two kinetic models admit the same solutions. This allows one, in particular, to adopt the well-studied numerical techniques for collisional kinetic equations, such as Monte Carlo algorithms, for the approximation of the solutions of the less standard jump process kinetic models. Next, we have moved to the problem of determining the asymptotic distribution functions, which depict the emerging aggregate trends of the system. For the jump process model there are not, in general, techniques allowing one to recover reliable approximations of the Maxwellian. Conversely, in the case of the collisional model one may take advantage of powerful asymptotic procedures, such as the quasi-invariant limit~\cite{toscani2006CMS}, which, for large times, transforms the collisional description into a mean field one expressed by a Fokker-Planck equation. The latter is often amenable to detailed analytical investigations, especially as far as the explicit determination of the Maxwellian is concerned. Thanks to the parallelism established before, we have shown that this asymptotic procedure may also be used to obtain approximate but explicit information on the steady distributions of the jump process model.

As concluding remarks, we would like to stress three relevant implications of the analysis performed in this paper.

First, the quasi-invariant limit described in Section~\ref{sect:quasi-invariant} may be used, in principle, to approximate the Maxwellian distribution produced by \textit{any} transition probability $T$, not just by those of the form~\eqref{eq:T.eps}. Indeed, as shown by~\eqref{eq:boltz.g-T}, the jump process model with~\eqref{eq:T.eps} is actually ruled by the transition probability $\cT$ on a suitably large time scale. Therefore, in order to explore the large time trend for a given $T$, it is sufficient to introduce fictitiously a small parameter $\epsilon>0$ and to define the quasi-invariant transition probability $\tilde{T}(v\,\vert\,\pr{v},\,\pr{v_\ast})=(1-\epsilon)\delta(v-\pr{v})+\epsilon T(v\,\vert\,\pr{v},\,\pr{v_\ast})$. At this point, the collision rule~\eqref{eq:binary_quasi-inv}, with $V_\cT$, $D_\cT$ replaced respectively by $V_T$, $D_T$, provides the basis for performing the quasi-invariant limit $\epsilon\to 0^+$ on the larger time scale $\tau=\epsilon t$, whence to deduce an approximation of the large time kinetic distribution produced by the microscopic $T$-dynamics and, in particular, of the Maxwellian. It is worth pointing out that, for a generic transition probability $T$, there are not, in many cases, other procedures able to yield more precise descriptions of the steady state of the jump process model.

Second, we observe that the scaled collision rule~\eqref{eq:binary_quasi-inv} implies, on the larger time scale $\tau=\epsilon t$, an evolution of both the mean and the energy which, for every $\epsilon>0$, is identical to that of the unscaled rule ($\epsilon=1$). This indicates that, starting from \textit{any} given collision rule~\eqref{eq:binary_gen}, the passage through the quasi-invariant transition probability~\eqref{eq:T.eps} may be used to obtain a quasi-invariant scaling which, on the $\tau$-scale, reproduces exactly the trends of the mean and also of the energy for every $\epsilon>0$. For this, it is formally sufficient to consider any transition probability $\cT$ such that $V_\cT=I$ and $D_\cT=D$. We stress that, while quasi-invariant scalings typically conserve the evolution of the mean for every $\epsilon>0$, the same is in general not true for the energy, whose equation normally contains a remainder, which vanishes only in the limit $\epsilon\to 0^+$, cf.~\cite{furioli2017M3AS}.

Third, as already quickly implied along the paper, the link between the jump process equation~\eqref{eq:boltz.T} and the collisional equation~\eqref{eq:boltz.strong} realised through the collision rules~\eqref{eq:binary} suggests that the collisional model may be used to obtain \textit{macroscopic} models from the microscopic jump process by means of the \textit{hydrodynamic limit}. The latter relies on the local equilibrium closure of the moment equations deduced by integrating~\eqref{eq:boltz.T} against suitable powers of the microscopic state $v$, for which the knowledge of the Maxwellian is necessary. Since the collisional model reproduces exactly the time evolution of the mean and of the energy of the jump process model, one may profitably use the Maxwellian obtained from the quasi-invariant limit of the collisional model to close the macroscopic equations derived in the hydrodynamic limit from the jump process model. Indeed, such a closure requires typically only the first two moments of the local equilibrium distribution. This may allow for more accurate \textit{hyperbolic} descriptions of the macroscopic dynamics, as opposed to the \textit{diffusive} ones typically recovered by means of the diffusive limit in the absence of sufficiently detailed information on the Maxwellian, cf.~\cite{hillen2000SIAP}. This argument also stresses that the jump process model and the collisional model are actually indistinguishable in terms of the macroscopic physics. In addition to that, the collisional model often offers more precise insights into the mesoscopic characteristics of the system.

A currently open problem, which may certainly deserve future attention, is the estimation of a suitable distance between the (often unknown) Maxwellian of the jump process model and the Fokker-Planck approximation of the Maxwellian of the collisional model. In particular, this could allow one to identify specific classes of transition probabilities $T$, hence of jump process models, for which such an approximation is actually \textit{a priori} reliable.

\section*{Acknowledgements}
This research was partially supported  by the Italian Ministry for Education, University and Research (MIUR) through the ``Dipartimenti di Eccellenza'' Programme (2018-2022), Department of Mathematical Sciences ``G. L. Lagrange'', Politecnico di Torino (CUP: E11G18000350001) and through the PRIN 2017 project (No. 2017KKJP4X) ``Innovative numerical methods for evolutionary partial differential equations and applications''.

This work is also part of the activities of the Starting Grant ``Attracting Excellent Professors'' funded by ``Compagnia di San Paolo'' (Torino) and promoted by Politecnico di Torino.

The PhD scholarship of NL is funded by ``Compagnia di San Paolo'' (Torino).

Both authors are members of GNFM (Gruppo Nazionale per la Fisica Matematica) of INdAM (Istituto Nazionale di Alta Matematica), Italy.

\appendix

\section{The Monte Carlo method for collisional kinetic equations}
\label{app:MC}
For the sake of completeness, in this appendix we quickly review some basic notions about the application of Monte Carlo algorithms to the approximate solution of collisional kinetic equations. The interested reader is referred to e.g.,~\cite{pareschi2001ESAIMP,pareschi2013BOOK} for more details and further references.

Let us consider the strong form of the collisional Boltzmann-type equation, cf.~\eqref{eq:boltz.strong}, which we may conveniently rewrite as
\begin{equation}
	\partial_tf=\frac{1}{\epsilon}\left(Q^+(f,\,f)-f\right),
	\label{eq:boltz.strong_Q+}
\end{equation}
where
$$ Q^+(f,\,f)(t,\,v):=\ave*{\int_\R\frac{1}{\pr{J}}f(t,\,\pr{v})f(t,\,\pr{v_\ast})\,dv_\ast}\geq 0 $$
is the so-called \textit{gain operator}. We have considered the coefficient $\frac{1}{\epsilon}$ in order to include in this discussion also the scaled equation~\eqref{eq:boltz.g}. Notice that for $\epsilon=1$ we obtain, in particular, precisely~\eqref{eq:boltz.strong}. Integrating both sides of~\eqref{eq:boltz.strong_Q+} with respect to $v\in\R$ and recalling the mass conservation property, we obtain
$$ 0=\int_\R Q^+(f,\,f)(t,\,v)\,dv-\int_\R f(t,\,v)\,dv=\int_\R Q^+(f,\,f)(t,\,v)\,dv-1, \qquad \forall\,t>0, $$
whence we deduce that $Q^+(f,\,f)(t,\,\cdot)$ is, at every time, a probability density. Discretising now~\eqref{eq:boltz.strong_Q+} in time with the forward Euler formula, we find
\begin{equation}
	f^{n+1}(v)=\left(1-\frac{\Delta{t}}{\epsilon}\right)f^n(v)+\frac{\Delta{t}}{\epsilon}Q^+(f^n,\,f^n)(v),
	\label{eq:boltz.fwd_Eul}
\end{equation}
where $\Delta{t}\in (0,\,\epsilon]$ is a fixed time step and $f^n(v)\approx f(t^n,\,v)$ with $t^n:=n\Delta{t}$. Since both $f^n$ and $Q^+(f^n,\,f^n)$ are probability densities and the right-hand side of~\eqref{eq:boltz.fwd_Eul} is a convex combination of them, also $f^{n+1}$ remains a probability density. The interpretation of~\eqref{eq:boltz.fwd_Eul} in terms of the underlying stochastic particle system is clear: during a time step, two randomly chosen particles update their microscopic states according to the collision-like rule~\eqref{eq:binary}, encoded in $Q^+$, with probability $\frac{\Delta{t}}{\epsilon}$. Alternatively, they do not interact with the complementary probability $1-\frac{\Delta{t}}{\epsilon}$, thereby leaving the distribution function unchanged. If we fix, in particular, $\Delta{t}=\epsilon$ then any two randomly chosen particles always interact.

Starting from these considerations, a Monte Carlo approach for the numerical approximation of the solution to~\eqref{eq:boltz.strong} can be described as detailed in Algorithm~\ref{alg:nanbu}, cf. also~\cite{bobylev2000PRE,nanbu1980JPSJ,pareschi2013BOOK}. In particular, as observed in Section~\ref{sect:comparisons}, cf. Remark~\ref{rem:MC_T}, the same algorithm may be used to approximate also the solution to the jump process model~\eqref{eq:boltz.T}, provided $\eta$, $\eta_\ast$ are chosen like in~\eqref{eq:eta_standard}. In this case, the sampling invoked in line~\ref{alg:sample_eta} of Algorithm~\ref{alg:nanbu} needs to take into account that the distribution of $\eta$, $\eta_\ast$ varies from pair to pair of interacting particles.

\begin{algorithm}[!t]
	\caption{Nanbu-like algorithm for~\eqref{eq:binary}-\eqref{eq:boltz.strong_Q+}, cf.~\cite{bobylev2000PRE,nanbu1980JPSJ,pareschi2013BOOK}}
	\begin{algorithmic}[1]
		\STATE fix $N>1$ (number of particles, even), $\Delta{t}=\epsilon$
		\STATE sample $N$ particles from the initial distribution $f^0$; let $\{v_i^0\}_{i=1}^{N}$ be their microscopic states
		\FOR{$n=0,\,1,\,2,\,\dots$}
			\STATE sample uniformly $\frac{N}{2}$ pairs of indexes $(i,\,j)$ with $i,\,j\in\{1,\,\dots,\,N\}$, $i\neq j$ and no repetition
			\FOR{every sampled pair $(i,\,j)$}
				\STATE \label{alg:sample_eta} sample a value of $\eta$, $\eta_\ast$ from their (common) distribution
				\STATE set
						$\begin{cases}
							v_i^{n+1}:=V_T(v_i^n,\,v_j^n)+D_T(v_i^n,\,v_j^n)\eta \\
							v_j^{n+1}:=V_T(v_j^n,\,v_i^n)+D_T(v_j^n,\,v_i^n)\eta_\ast,
						\end{cases}$
						(cf.~\eqref{eq:binary})
			\ENDFOR
			\STATE \label{alg:hist} construct an approximation of $f^{n+1}$ from the samples $\{v_i^{n+1}\}_{n=1}^{N}$
		\ENDFOR
	\end{algorithmic}
	\label{alg:nanbu}
\end{algorithm}

\medskip

In the case of the consensus/dissensus dynamics, cf. Section~\ref{sect:Maxwellian}, the numerical solutions of the collisional model~\eqref{eq:boltz.strong}-\eqref{eq:binary.opdyn} and of the jump process model~\eqref{eq:boltz.g-T}-\eqref{eq:T.eps}, cf. Figure~\ref{fig:ginf.3-num}, have been computed by means of Algorithm~\ref{alg:nanbu} with $N=10^6$ particles. The number of iterations of the algorithm has been taken sufficiently large, so as to ensure that the steady state was reached within a resonable numerical tolerance. The $L^\infty$ relative error between the exact and the numerically computed mean and energy of both models is $O(10^{-3})$. Also the $L^\infty$ relative error between the analytical Maxwellian~\eqref{eq:ginf.3} and the large time numerical solution of the collisional model~\eqref{eq:boltz.strong}-\eqref{eq:binary.opdyn} is $O(10^{-3})$. Finally, the $L^\infty$ relative error between the large time solution of the collisional model and that of the jump process model is $O(10^{-2})$.

In the case of the random onset of Alzheimer's disease in the brain, cf. Section~\ref{sect:bounded_domains}, the transient solutions of the collisional model~\eqref{eq:boltz.strong}-\eqref{eq:alzheimer_collision} and of the jump process model~\eqref{eq:alzheimer_T}-\eqref{eq:alzheimer_jpke}, as well as their mean and energy, cf. Figures~\ref{fig:alzheimer_M1M2},~\ref{fig:alzheimer_evo}, have been computed numerically by means of Algorithm~\ref{alg:nanbu} with $N=10^6$ particles. Furthermore, the reconstruction of $f^n$, cf. line~\ref{alg:hist} of Algorithm~\ref{alg:nanbu}, has been performed by computing, at each time step, the normalised histogram of the microscopic states $\{v_i^n\}_{i=1}^{N}$ with $10^3$ uniform bins in the interval $[0,\,1]$.

\bibliographystyle{plain}
\bibliography{LnTa-kinetic_transprob}

\begin{thebibliography}{10}

\bibitem{agnelli2015M3AS}
J.~P. Agnelli, F.~Colasuonno, and D.~Knopoff.
\newblock A kinetic theory approach to the dynamics of crowd evacuation from
  bounded domains.
\newblock {\em Math. Models Methods Appl. Sci.}, 25(1):109--129, 2015.

\bibitem{albi2016SIAP}
G.~Albi, M.~Bongini, E.~Cristiani, and D.~Kalise.
\newblock Invisible control of self-organizing agents leaving unknown
  environments.
\newblock {\em SIAM J. Appl. Math.}, 76(4):1683--1710, 2016.

\bibitem{albi2015CMS}
G.~Albi, M.~Herty, and L.~Pareschi.
\newblock Kinetic description of optimal control problems and application to
  opinion consensus.
\newblock {\em Commun. Math. Sci.}, 13(6):1407--1429, 2015.

\bibitem{albi2014PTRSA}
G.~Albi, L.~Pareschi, and M.~Zanella.
\newblock Boltzmann-type control of opinion consensus through leaders.
\newblock {\em Phil. Trans. R. Soc. A}, 372(2028):20140138/1--18, 2014.

\bibitem{alt1980JMB}
W.~Alt.
\newblock Biased random walk models for chemotaxis and related diffusion
  approximations.
\newblock {\em J. Math. Biol.}, 9(2):147--177, 1980.

\bibitem{ambrosio2008BOOK}
L.~Ambrosio, N.~Gigli, and G.~Savar{\'e}.
\newblock {\em Gradient flows in metric spaces and in the space of probability
  measures}.
\newblock Lectures in Mathematics ETH Z\"urich. Birkh\"auser Verlag, Basel,
  2008.

\bibitem{bertotti2008MCM}
M.~L. Bertotti and M.~Delitala.
\newblock On a discrete generalized kinetic approach for modelling persuader's
  influence in opinion formation processes.
\newblock {\em Math. Comput. Modelling}, 48(7-8):1107--1121, 2008.

\bibitem{bertsch2017MMB}
M.~Bertsch, B.~Franchi, N.~Marcello, M.~C. Tesi, and A.~Tosin.
\newblock Alzheimer's disease: a mathematical model for onset and progression.
\newblock {\em Math. Med. Biol.}, 34(2):193--214, 2017.

\bibitem{bertsch2017JPA}
M.~Bertsch, B.~Franchi, M.~C. Tesi, and A.~Tosin.
\newblock Microscopic and macroscopic models for the onset and progression of
  {A}lzheimer's disease.
\newblock {\em J. Phys. A: Math. Theor.}, 50(41):414003/1--22, 2017.

\bibitem{bertsch2018SIMA}
M.~Bertsch, B.~Franchi, M.~C. Tesi, and A.~Tosin.
\newblock Well-posedness of a mathematical model for {A}lzheimer's disease.
\newblock {\em SIAM J. Math. Anal.}, 50(3):2362--2388, 2018.

\bibitem{bobylev2000PRE}
A.~V. Bobylev and K.~Nanbu.
\newblock Theory of collision algorithms for gases and plasmas based on the
  {B}oltzmann equation and the {L}andau-{F}okker-{P}lanck equation.
\newblock {\em Phys. Rev. E}, 61(4):4576--4586, 2000.

\bibitem{boltzmann1970CHAPTER}
L.~Boltzmann.
\newblock Weitere {S}tudien \"{u}ber das {W}\"{a}rmegleichgewicht unter
  {G}asmolek\"{u}len.
\newblock In {\em Kinetische Theorie II. WTB Wissenschaftliche
  Taschenb\"{u}cher}. Vieweg+Teubner Verlag, Wiesbaden, 1970.

\bibitem{boudin2009M2NA}
L.~Boudin and F.~Salvarani.
\newblock A kinetic approach to the study of opinion formation.
\newblock {\em ESAIM Math. Model. Numer. Anal.}, 43(3):507--522, 2009.

\bibitem{chauviere2007NHM}
A.~Chauvi{\`e}re, T.~Hillen, and L.~Preziosi.
\newblock Modeling cell movement in anisotropic and heterogeneous network
  tissues.
\newblock {\em Netw. Heterog. Media}, 2(2):333--357, 2007.

\bibitem{cordier2005JSP}
S.~Cordier, L.~Pareschi, and G.~Toscani.
\newblock On a kinetic model for a simple market economy.
\newblock {\em J. Stat. Phys.}, 120(1):253--277, 2005.

\bibitem{delitala2014KRM}
M.~Delitala and T.~Lorenzi.
\newblock A mathematical model for value estimation with public information and
  herding.
\newblock {\em Kinet. Relat. Models}, 7(1):29--44, 2014.

\bibitem{dolfin2017KRM}
M.~Dolfin, D.~Knopoff, L.~Leonida, and D.~Maimone Ansaldo~Patti.
\newblock Escaping the trap of `blocking': {A} kinetic model linking economic
  development and political competition.
\newblock {\em Kinet. Relat. Models}, 10(2):423--443, 2017.

\bibitem{duering2018EPJB}
B.~D\"{u}ring, L.~Pareschi, and G.~Toscani.
\newblock Kinetic models for optimal control of wealth inequalities.
\newblock {\em Eur. Phys. J. B}, 91:265/1--12, 2018.

\bibitem{festa2018KRM}
A.~Festa, A.~Tosin, and M.-T. Wolfram.
\newblock Kinetic description of collision avoidance in pedestrian crowds by
  sidestepping.
\newblock {\em Kinet. Relat. Models}, 11(3):491--520, 2018.

\bibitem{freguglia2017CMS}
P.~Freguglia and A.~Tosin.
\newblock Proposal of a risk model for vehicular traffic: {A} {B}oltzmann-type
  kinetic approach.
\newblock {\em Commun. Math. Sci.}, 15(1):213--236, 2017.

\bibitem{furioli2017M3AS}
G.~Furioli, A.~Pulvirenti, E.~Terraneo, and G.~Toscani.
\newblock {F}okker-{P}lanck equations in the modeling of socio-economic
  phenomena.
\newblock {\em Math. Models Methods Appl. Sci.}, 27(1):115--158, 2017.

\bibitem{gualandi2018M3AS}
S.~Gualandi and G.~Toscani.
\newblock Call center service times are lognormal: {A} {F}okker-{P}lanck
  description.
\newblock {\em Math. Models Methods Appl. Sci.}, 28(8):1513--1527, 2018.

\bibitem{gualandi2019M3AS}
S.~Gualandi and G.~Toscani.
\newblock Human behavior and lognormal distribution. a kinetic description.
\newblock {\em Math. Models Methods Appl. Sci.}, 29(4):717--753, 2019.

\bibitem{gualandi2019PHYSA}
S.~Gualandi and G.~Toscani.
\newblock Size distribution of cities: {A} kinetic explanation.
\newblock {\em Phys. A}, 524:221--234, 2019.

\bibitem{hillen2006JMB}
T.~Hillen.
\newblock {$M^5$} mesoscopic and macroscopic models for mesenchymal motion.
\newblock {\em J. Math. Biol.}, 53(4):585--616, 2006.

\bibitem{hillen2000SIAP}
T.~Hillen and H.~G. Othmer.
\newblock The diffusion limit of transport equations derived from velocity-jump
  processes.
\newblock {\em SIAM J. Appl. Math.}, 61:751--775, 2000.

\bibitem{klar1997JSP}
A.~Klar and R.~Wegener.
\newblock Enskog-like kinetic models for vehicular traffic.
\newblock {\em J. Stat. Phys.}, 87(1-2):91--114, 1997.

\bibitem{loy2019JMB}
N.~Loy and L.~Preziosi.
\newblock Kinetic models with non-local sensing determining cell polarization
  and speed according to independent cues.
\newblock {\em J. Math. Biol.}, 2019.
\newblock To appear.

\bibitem{loy2019PREPRINT}
N.~Loy and M.~Zanella.
\newblock Structure preserving schemes for nonlinear {F}okker-{P}lanck
  equations with anisotropic diffusion.
\newblock Preprint (\texttt{doi:10.13140/RG.2.2.11832.39685}), 2019.

\bibitem{nanbu1980JPSJ}
K.~Nanbu.
\newblock Direct simulation scheme derived from the {B}oltzmann equation. {I}.
  {M}onocomponent gases.
\newblock {\em J. Phys. Soc. Japan}, 49(5):2042--2049, 1980.

\bibitem{pareschi2001ESAIMP}
L.~Pareschi and G.~Russo.
\newblock An introduction to {M}onte {C}arlo method for the {B}oltzmann
  equation.
\newblock {\em ESAIM: Proc.}, 10:35--75, 2001.

\bibitem{pareschi2013BOOK}
L.~Pareschi and G.~Toscani.
\newblock {\em Interacting {M}ultiagent {S}ystems: {K}inetic equations and
  {M}onte {C}arlo methods}.
\newblock Oxford University Press, 2013.

\bibitem{pareschi2017JSC}
L.~Pareschi and M.~Zanella.
\newblock Structure preserving schemes for nonlinear {F}okker-{P}lanck
  equations and applications.
\newblock {\em J. Sci. Comput.}, 2017.

\bibitem{prigogine1960OR}
I.~Prigogine and F.~C. Andrews.
\newblock A {B}oltzmann-like approach for traffic flow.
\newblock {\em Operations Res.}, 8(6):789--797, 1960.

\bibitem{prigogine1971BOOK}
I.~Prigogine and R.~Herman.
\newblock {\em Kinetic theory of vehicular traffic}.
\newblock American Elsevier Publishing Co., New York, 1971.

\bibitem{puppo2017CMS}
G.~Puppo, M.~Semplice, A.~Tosin, and G.~Visconti.
\newblock Analysis of a multi-population kinetic model for traffic flow.
\newblock {\em Commun. Math. Sci.}, 15(2):379--412, 2017.

\bibitem{puppo2017KRM}
G.~Puppo, M.~Semplice, A.~Tosin, and G.~Visconti.
\newblock Kinetic models for traffic flow resulting in a reduced space of
  microscopic velocities.
\newblock {\em Kinet. Relat. Models}, 10(3):823--854, 2017.

\bibitem{stroock1974ZWVG}
D.~W. Stroock.
\newblock Some stochastic processes which arise from a model of the motion of a
  bacterium.
\newblock {\em Z. Wahrsch. Verw. Gebiete}, 28(4):305--315, 1974.

\bibitem{toscani2006CMS}
G.~Toscani.
\newblock Kinetic models of opinion formation.
\newblock {\em Commun. Math. Sci.}, 4(3):481--496, 2006.

\bibitem{toscani2018PRE}
G.~Toscani, A.~Tosin, and M.~Zanella.
\newblock Opinion modeling on social media and marketing aspects.
\newblock {\em Phys. Rev. E}, 98(2):022315/1--15, 2018.

\bibitem{tosin2019MMS}
A.~Tosin and M.~Zanella.
\newblock Kinetic-controlled hydrodynamics for traffic models with
  driver-assist vehicles.
\newblock {\em Multiscale Model. Simul.}, 17(2):716--749, 2019.

\bibitem{tosin2019MCRF_preprint}
A.~Tosin and M.~Zanella.
\newblock Uncertainty damping in kinetic traffic models by driver-assist
  controls.
\newblock Preprint (\texttt{doi:10.13140/RG.2.2.35871.41124}), 2019.

\bibitem{villani1998PhD}
C.~Villani.
\newblock {\em Contribution \`{a} l'\'{e}tude math\'{e}matique des
  \'{e}quations de Boltzmann et de Landau en th\'{e}orie cin\'{e}tique des gaz
  et des plasmas}.
\newblock Ph{D} thesis, Paris 9, 1998.

\bibitem{villani1998ARMA}
C.~Villani.
\newblock On a new class of weak solutions to the spatially homogeneous
  {B}oltzmann and {L}andau equations.
\newblock {\em Arch. Ration. Mech. Anal.}, 143(3):273--307, 1998.

\end{thebibliography}

\end{document}